\documentclass[structuredabstrac]{aa}  

\usepackage{graphicx}
\usepackage{natbib}
\usepackage{xspace}
\xspaceaddexceptions{]\}}

\usepackage[update,prepend]{epstopdf}
\usepackage{booktabs}
\usepackage{amsmath}

\newcommand{\rarrow}{\ensuremath{\rightarrow}}

\newcommand{\CCr}{$N$(\CthHtw)/$N$(\CCH)\xspace}

\newcommand{\GILDAS}{\texttt{GILDAS}\xspace}
\newcommand{\CLASS}{\texttt{CLASS}\xspace}

\newcommand{\IRAMthm}{\textrm{IRAM-30m}\xspace}

\newcommand{\herschel}{\textit{Herschel}\xspace}

\newcommand{\Cp}{\ion{C}{ii}\xspace}

\newcommand{\HII}{\ion{H}{ii}\xspace}

\newcommand{\HH}    {\mbox{H$_2$}\xspace}           % H2
\newcommand{\thCO}  {\mbox{$^{13}$CO}\xspace}       % 13CO
\newcommand{\CeiO}  {\mbox{C$^{18}$O}\xspace}       % C18O
\newcommand{\CH}  {\mbox{CH}\xspace}              % CCH
\newcommand{\CHp}  {\mbox{CH$^+$}\xspace}              % CCH
\newcommand{\CCH}  {\mbox{CCH}\xspace}              % CCH
\newcommand{\CthHtw}  {\mbox{c-C$_3$H$_2$}\xspace}  % c-C3H2
\newcommand{\CfoH}  {\mbox{C$_4$H}\xspace}  % C4H
\newcommand{\emm}[1]{\ensuremath{#1}}   % Ensures math mode.
\newcommand{\emr}[1]{\emm{\mathrm{#1}}} % Uses math roman fonts.
\newcommand{\unit}[1]{\emm{\, \emr{#1}}}
\newcommand{\K}   {\unit{K}\xspace}

\newcommand{\mum} {\unit{\mu m}\xspace}
\newcommand{\cmq}{\unit{cm^{-2}}\xspace}
\newcommand{\cmc}{\unit{cm^{-3}}\xspace}

\newcommand{\kms}   {\unit{km\,s^{-1}}\xspace}
\newcommand{\Kkms}{\unit{K\,km\,s^{-1}}\xspace}

\newcommand{\GHz} {\unit{GHz}\xspace}

\newcommand{\Av}{\emm{A_V}\xspace}

\newcommand{\hab}{\emm{G_{\rm 0}}\xspace}
\newcommand{\Tas}{\emm{T_\emr{A}^*}\xspace}
\newcommand{\Tmb}{\emm{T_\emr{mb}}\xspace}

\newcommand{\Trot}{\emm{T_\emr{rot}}\xspace}
\newcommand{\Beff}{\emm{B_\emr{eff}}\xspace}
\newcommand{\Feff}{\emm{F_\emr{eff}}\xspace}
\newcommand{\Tkin}{\emm{T_\emr{kin}}\xspace}

\newcommand{\nhh}{\emm{n_\emr{H_2}}\xspace}

\newcommand{\ex}[2]{\ensuremath{#1 \times 10^{#2}}\xspace}

\usepackage{txfonts}
\usepackage{ulem}

\newcommand{\stout}[1]{\textbf{{\color{red}\sout{#1}}}}
\renewcommand{\stout}[1]{}

\begin{document}

\title{ Spatial distribution of small hydrocarbons in the neighborhood of the Ultra Compact HII region Monoceros~R2\thanks{Herschel is an ESA space observatory with science instruments provided by European-led Principal Investigator consortia and with important participation from NASA.}\fnmsep\thanks{Based on observations carried out with the IRAM 30m Telescope. IRAM is supported by INSU/CNRS (France), MPG (Germany) and IGN (Spain).} }
%\subtitle{More evidence for the link with PAHs}

\author{
P.~Pilleri\inst{1,2,3},
S.~Trevi\~no-Morales\inst{4},
A.~Fuente\inst{2}, 
C.~Joblin\inst{5,6},
J.~Cernicharo\inst{1},
M.~Gerin\inst{7},
S.~Viti\inst{8},
O.~Bern\'e\inst{5,6},
J.R.~Goicoechea\inst{1},
J.~Pety\inst{9},
M.~Gonzalez-Garc\'ia\inst{4},
J.~Montillaud\inst{10},
V.~Ossenkopf\inst{11},
C.~Kramer\inst{4},
S.~Garc{\'i}a-Burillo\inst{2},
F.~Le~Petit\inst{12},
J.~Le~Bourlot\inst{12}
}

 \institute{%
 %1
Centro de Astrobiolog\'{\i}a (INTA-CSIC),
             Ctra. M-108, km.~4, E-28850 Torrej\'on de Ardoz, Spain
\and
%2
Observatorio Astron\'omico Nacional, Apdo. 112, E-28803 Alcal\'a de Henares (Madrid), Spain
\and
Los Alamos National Laboratory, Los Alamos, NM 87545, USA
\and
%3
Instituto de Radio Astronom\'ia Milim\'etrica (IRAM), Avenida Divina Pastora 7, Local 20, 18012 Granada, Spain
%4
\and
Universit\'e de Toulouse; UPS-OMP; IRAP;  Toulouse, France
\and
%5
CNRS; IRAP; 9 Av. colonel Roche, BP 44346, F-31028 Toulouse cedex 4, France 
\and
LERMA, Observatoire de Paris, 61 Av. de l'Observatoire, 75014 Paris, France 
\and
%6
Dept. of Physics and Astronomy, UCL, Gower Place, London WC1E6BT, UK
\and
%7
Institut de Radioastronomie Millim\'etrique, 300 Rue de la Piscine, 38406 Saint Martin d'H\'eres, France
 \and
 %8
Department of Physics, P.O.Box 64, FI-00014, University of Helsinki, Finland
\and
%10
 I. Physikalisches Institut der Universit\"at zu K\"oln, Z\"ulpicher Stra\ss{}e 77, 50937 K\"oln, Germany
\and
%9
Observatoire de Paris, LUTH and Universit\'e Denis Diderot, Place J.
Janssen, 92190 Meudon, France
}

\authorrunning {P. Pilleri, et al.} 
\titlerunning{Spatial distribution of \CCH and \CthHtw in the Ultra Compact HII region Monoceros~R2}

%==========================================================================================
%==========================================================================================
% ABSTRACT
%==========================================================================================
%==========================================================================================

\abstract
% context heading (optional)
{
We study the chemistry of small hydrocarbons in the photon-dominated regions (PDRs) associated with the ultra-compact \HII region  (UC\HII) Mon~R2.
}
{
Our goal is to determine the variations of the abundance of small hydrocarbons in a high-UV irradiated PDR and investigate the chemistry of these species.
}
% methods heading (mandatory)
{
We present an observational  study of the small hydrocarbons \CH, \CCH and \CthHtw in Mon~R2 combining spectral mapping data obtained with the \IRAMthm telescope and the \herschel space observatory. We determine the column densities of these species, and compare their spatial distributions with that of polycyclic aromatic hydrocarbon (PAH), which trace the PDR. We compare the observational results with different chemical models  to explore the relative importance of gas-phase, grain-surface and time-dependent chemistry in these environments.} 
% results heading (mandatory)
{The emission of the small hydrocarbons show  different spatial patterns. The 
\CCH emission is  extended while \CH and   \CthHtw are concentrated towards the more illuminated layers of the PDR. The ratio of the column densities of  
\CthHtw and \CCH shows spatial variations up to a factor of a few, increasing 
from $N(\CthHtw)/N(\CCH)\approx0.004$ in the envelope to a
maximum of $\approx0.015-0.029$
towards the 8\,$\mu$m emission peak.
Comparing these results with other galactic PDRs, 
we find that the abundance of \CCH is quite constant over a wide range of \hab, 
whereas the abundance of \CthHtw is higher in low-UV PDRs, with the 
$N(\CthHtw)/N(\CCH)$ ratio ranging $\approx$0.008-0.08 from high to low UV PDRs.
In Mon~R2, the gas-phase steady-state chemistry can account 
relatively well for the abundances of \CH and \CCH in the most exposed layers of 
the PDR, but falls short by a factor of 10 to reproduce \CthHtw.    
 In the low-density molecular envelope, time-dependent effects and grain surface
chemistry play a dominant role in determining the hydrocarbons abundances. }
{Our study shows that the small hydrocarbons \CCH and \CthHtw present a complex chemistry in which UV photons, grain-surface 
chemistry and time dependent effects contribute to determine their abundances. Each of these effects may be dominant depending on the local physical conditions, and the superposition of different regions along the  line of sight leads to
 the variety of measured abundances. }


\maketitle
%==========================================================================================
%==========================================================================================
% INTRO
%==========================================================================================
%==========================================================================================

\section{Introduction}

Ultracompact \HII regions (UC\HII)  represent one of the earliest phases in the formation of a massive star.  In these environments, the chemistry  is strongly driven by the mutual interaction between the gas, the dust and the UV photons from nearby stars. The H-ionizing photons are  absorbed in a thin layer around the star, forming the \HII region, while the UV radiation carrying energies less than 13.6~eV penetrate in the 
deeper layers of the molecular cloud and produce  the so-called  photon-dominated region (PDR). In these environments, the thermal balance, the chemistry and ionization balance are driven by UV photons. 
PDRs associated  with UC\HII are characterized by extreme UV field intensities \citep[expressed in units of the Habing field \hab, see][]{habing68} and small physical scales ($\lesssim0.1$\,pc). These environments are therefore ideal to study the effect of a strong UV field  on the chemistry of gas-phase species. 

The chemistry of small hydrocarbons in PDRs is still poorly constrained. 
The comparison of observations and gas-phase chemical models showed that this type of chemistry alone cannot account for the abundance of small hydrocarbons, especially for those species containing more than two C atoms: in NGC 7023, the Orion Bar \citep{fuente03}, IC63 and the Horsehead nebula \citep{teyssier04} the abundances  of species such as  \CthHtw and \CfoH derived from the observations are at least an order of magnitude larger compared to model predictions. 
This discrepancy was confirmed by high spatial resolution observations and detailed modeling of the Horsehead nebula \citep{pety05}, in which the emission of small hydrocarbons was compared to other PDR tracers such as the mid-IR emission in the aromatic infrared bands (AIBs) due to polycyclic aromatic hydrocarbons (PAHs). The authors suggested that small hydrocarbons may form through non gas-phase processes such as  the photo-destruction of the AIB carriers.  
 In the Horsehead \citep{pety05}  and in NGC~7023 (Pilleri et al., in prep.), the spatial distributions of \CCH and \CthHtw\  are strikingly similar.    Similarly, a tight correlation between these two hydrocarbons has been observed by \citet{lucas00} and \citet{gerin11}  on several lines of sights in the diffuse medium.  The small radical \CH has also a relatively constant abundance in these environments \citep{sheffer08}. In  low- to mild-UV irradiated PDRs ($\hab \sim 100$ for the Horsehead and $\sim 2600$ for NGC7023) the tight correlation between \CCH and \CthHtw suggests a common origin of these two small hydrocarbons.  However, this spatial correlation has not been explored yet in highly UV-illuminated PDRs ($\hab \gtrsim \ex{1}{4}$).

In this paper, we present new observations of several lines of \CH, \CCH and \CthHtw in  the PDRs associated with the UC\HII region Mon~R2 and investigate observationally the relative variations of these species.  In Sect.\,\ref{sec_observations} we present the source and the observations, and the results are reported in Sect.\,\ref{sec_results}. In Sect.\,\ref{subsec_coldens} we  calculate  column densities and abundances, while in Sect.\,\ref{sec_discussion} we compare the spatial distributions to that of other PDR tracers and discuss the quantitative consistence with different chemical models. Finally, Sect.\,\ref{sec_conclusions} presents the conclusions and perspectives.

\section{Observations}

\label{sec_observations}

\subsection{Mon~R2}
\begin{figure}
\centering
\includegraphics[width=0.9\linewidth]{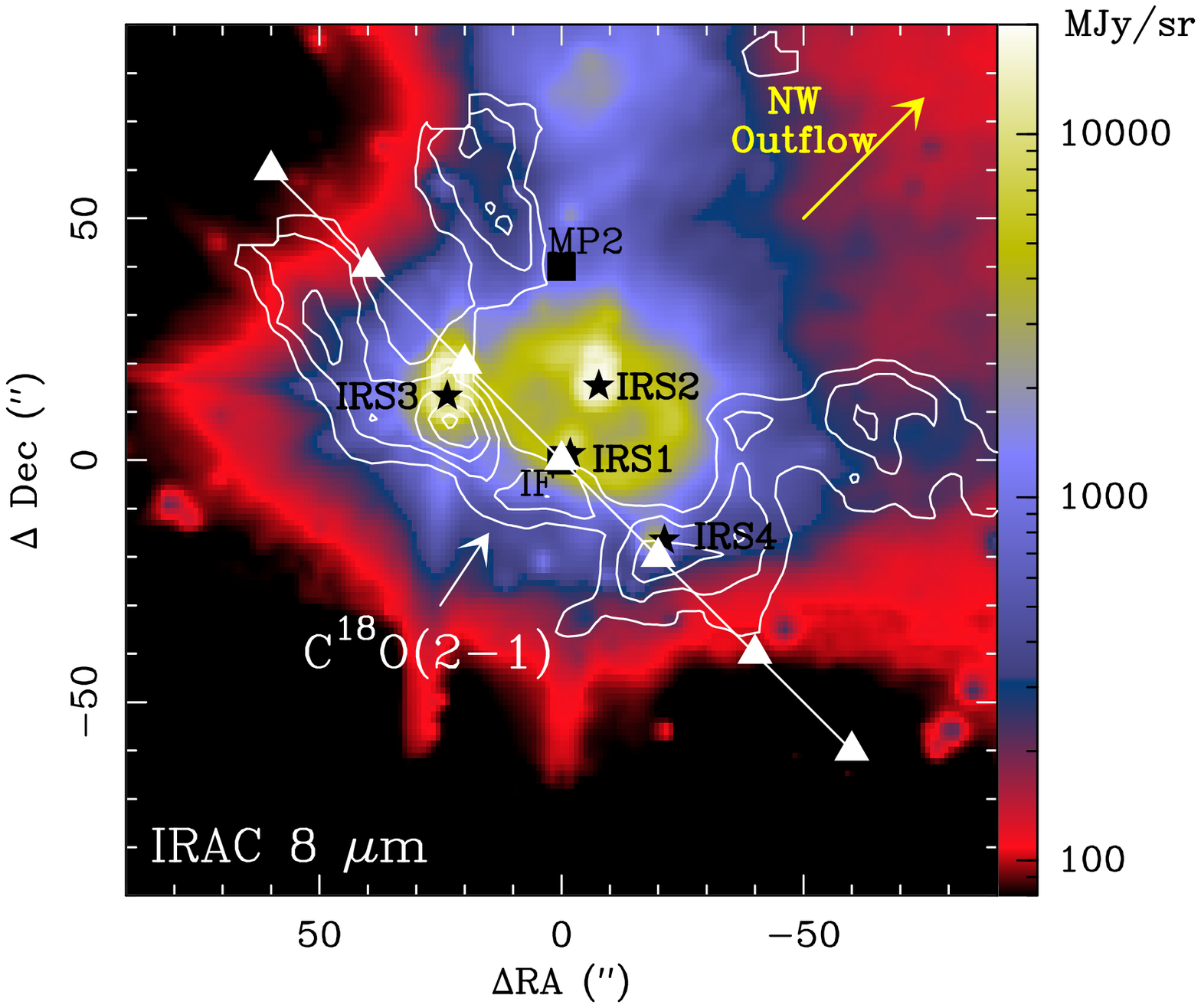}
\caption{Overview of the region. The color image displays the IRAC 8\mum emission \citep{ginard12}, in which both the bright circular PDR and the 
extended northern PDR can be distinguished beneath the peaks due to the young stellar objects. The white contours represent the integrated intensity 
of the \CeiO (2\rarrow1) line  between 5 and 15~\kms observed 
at the IRAM 30m telescope \citep{pilleri12a}, with contour levels  from 16 to 24\K\kms in linear steps of 2\Kkms. 
The black squares represent 
the positions of the  ionization front (IF) and the molecular peak (MP2), whereas the triangles and the white line indicate the positions observed with HIFI.  The black stars indicate the position of the most intense infrared sources following the nomenclature of  \citet{henning92}.  The yellow arrow indicates the direction of the NW molecular outflow \citep{tafalla94}.
}
\label{fig_mapmon2}
\end{figure}

Mon~R2 is a close-by ultracompact \HII region (d=830 pc), which comprises several PDRs that can be spatially resolved in both the mm and IR domains. Due to its brightness and proximity, this source can be used as reference for other PDRs illuminated by a strong UV field.
 The most intense UV source is called IRS1 and it is located at the center of the  UC\HII (see Fig.\ref{fig_mapmon2}). A  low-density molecular cloud surrounds this innermost region and extends for several arcminutes. The brightest PDR of Mon~R2, which surrounds the UC\HII and peaks at about 20\arcsec\, to the NW of IRS1, shows  both very intense AIB emission and \HH rotational lines \citep{berne09a} and is due to the illumination of the internal walls of the molecular cloud by IRS1. This PDR is very compact ($d \sim 5\arcsec$) and illuminated by a very strong radiation field ($\hab \sim \ex{5}{5}$). 
A secondary PDR, detected north of the UC\HII at a distance of about 1\arcmin\ from IRS1, is much fainter and more extended. The northern Molecular Peak position at [0\arcsec\,;40\arcsec]  \citep[hereafter MP2  following the nomenclature of][]{ginard12} is clearly detected in the AIB emission at 8\mum, and its chemical properties seem to be similar to those of low- to mild-UV irradiated PDRs \citep{ginard12}. The origin of this PDR could be the outer envelope of the molecular cloud illuminated by an external UV field, or it could be due to UV photons that are escaping from IRS1 through a hole, illuminating its walls.

\begin{figure}[ht]
\centering
\includegraphics[width=1\linewidth]{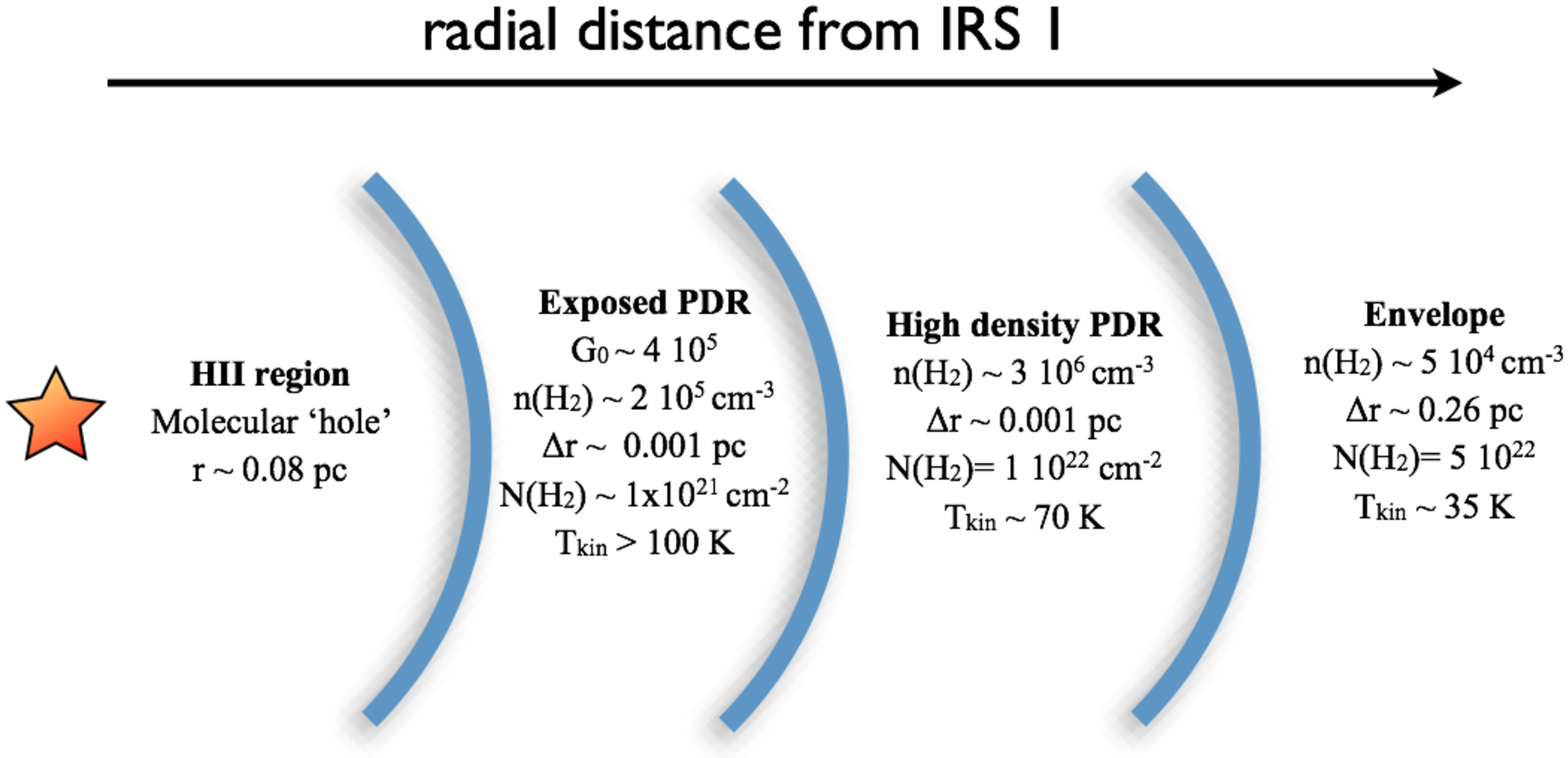}
\caption{A 1D schematic of the geometry of MonR2 as a function of the radial distance to IRS1. } 
\label{fig_schema}
\end{figure}

Although there are some asymmetries in the general shape of the cloud, such as the 
molecular hole to the NW, we can assume  as a zero-order approximation that the
dense core around the UC\HII region is spherical and composed by concentric 
shells centered on IRS1, a simplified model which is close to reality in most 
outwards directions.
Each of the shells has its own physical conditions (kinetic temperature, gas density, \hab) which vary with radius. In the following, we will assume the physical parameters presented in \citet{pilleri12a} and schematized in Fig.\,\ref{fig_schema}: the region immediately surrounding the star is an \HII region, which is expanding at a velocity of $\sim 10$\kms \citep{choi00} and which is free of any molecular gas. The diameter of the region is $\sim 40$\arcsec, corresponding to $\sim 0.08$\,pc. This is followed by a thin (1~mag), high UV-irradiated PDR, which has a density   of $\nhh\sim\ex{2}{5}$\cmc and a gas kinetic temperature of $\sim100$\K   \citep{berne09a}. Going farther from IRS1, there is a thicker ($\sim 8~$mag), warm ($\Tkin\sim70$\K) and high density \citep[\nhh$\sim\ex{3}{6}$\cmc, ][]{rizzo03} PDR, which is expanding at a velocity of $\lesssim 1$\kms \citep{fuente10,pilleri12a}.  
This  high-density layer is surrounded by the parent molecular cloud, which is about 40  mag thick, has a relatively low density (\nhh$\sim5\times 10^4$\cmc) and temperature (\Tkin$\sim 35\K$), and is rather quiescent \citep{fuente10, pilleri12a}.  The cloud is also illuminated from the outside with a UV field of $\hab \sim 100$. 

The presence of these different phases and its proximity make Mon~R2 the best source to investigate the chemistry and physics of these extreme PDRs and their parent molecular clouds.

%%%%%%%%%%%%%%%%%%%%%%%%%%%%%%%%%%%%%%%%
%%%%%%%%%%%%%%%%%%%%%%%%%%%%%%%%%%%%%%%%
%%%%%%%%%%%%%%%%%%%%%%%%%%%%%%%%%%%%%%%%
%%%%%%%%%%%%%%%%%%%%%%%%%%%%%%%%%%%%%%%%
%%%%%%%%%%%%%%%%%%%%%%%%%%%%%%%%%%%%%%%%
%%%%%%%%%%%%%%%%%%%%%%%%%%%%%%%%%%%%%%%%
%%%%%%%%%%%%%%%%%%%%%%%%%%%%%%%%%%%%%%%%

\begin{table*}
\caption{Summary of the observations}
\label{table_obssum}
\begin{center}
\begin{tabular}{llcccccc}
\toprule
Line(s) 	& 	Frequency   & Instrument  & FWHM  & $\eta_l$ \tablefootmark{a}  & Position      & OM\tablefootmark{b}	 & $T_{\rm{rms}}$\tablefootmark{c} 	\\
		&	[\GHz]  &	& ["]   & 	& 		& 		& [\K]	\\
\midrule
C$^{18}$O 	$J = 2-1$  & 219.560   & 	IRAM-HERA 	& 	10.0 		& 0.67	& Map & OTF &     0.025  \\
C$^{18}$O 	$J =5-4$  	& 548.830   &	 \herschel-HIFI & 	 38.6 	 & 0.79   	& Strip & OTF &    0.008    \\	
\midrule
\CH $J = 3/2 - 1/2$		& 	536.761 \tablefootmark{*} & \herschel-HIFI	& 38  & 0.77  & Strip &OTF  	& 0.041   \\
\CH $J = 5/2 - 3/2$		& 	1656.961 \tablefootmark{*} & \herschel-HIFI	& 12.8  & 0.74  & IF &SP	& 0.284  \\
\midrule
\CCH  $N = 1-0$ 	& 	87.317 \tablefootmark{*}  & 	IRAM 		& 29		&0.85 & Map	& OTF    & 0.070\\
\CCH  $N = 3-2$ 	& 	262.004 \tablefootmark{*}  	 & 	IRAM-HERA 		& 9	&0.6 & Map		& OTF    & 0.380\\
\CCH  $N = 6-5$ 	& 	523.972 \tablefootmark{*}  	 & 	\herschel-HIFI		& 40.5	&0.78 & Strip	& OTF    &0.057 \\
\midrule
\CthHtw $J_{K_A,K_C} = 2_{1,2}-1_{0,1}$	& 	85.33889		&	IRAM	& 29		& 0.85	&Map		&OTF		 & 0.070 \\
\CthHtw $J_{K_A,K_C} = 4_{0,4}-3_{1,3}$	& 	150.82067	&	IRAM	& 16		& 0.64	&Map		& OTF	 & 0.054 \\
\CthHtw $J_{K_A,K_C} = 4_{1,4}-3_{0,3}$	& 	150.85191	&	IRAM	& 16		& 0.64	& Map	& OTF		 & 0.054 \\
\CthHtw $J_{K_A,K_C} = 6_{1,6}-5_{0,5}$	& 	217.82215\tablefootmark{*} 	&	IRAM-HERA	& 11		& 0.61	& Map	& OTF	& 0.016\\
\CthHtw $J_{K_A,K_C} = 5_{1,4}-4_{2,3}$		&  217.94005 	&	IRAM	& 11	& 0.61	& Map	& OTF	& 0.150\\
\CthHtw $J_{K_A,K_C} = 4_{3,2}-4_{2,1}$		&  227.16913 	&	IRAM	& 10	& 0.59	& Map	& OTF	& 0.150\\
\CthHtw $J_{K_A,K_C} = 7_{0,7}-6_{1,6}$		&  251.31434\tablefootmark{*} 	&	IRAM	& 9	& 0.55	& Map	& OTF	& 0.115\\
\CthHtw $J_{K_A,K_C} = 6_{2,5}-5_{1,6}$		&  251.52730	&	IRAM	& 9	& 0.55	& Map	& OTF	& 0.115\\
\bottomrule
\end{tabular}
\end{center}
\tablefoottext{a}{$\eta_l = B_{eff}/F_{eff}$}.\\
\tablefoottext{b}{Observing Modes: on the fly (OTF), single pointing (SP).}\\
\tablefoottext{c}{Calculated on a \Tmb scale with $\Delta v = 0.5\kms$.}\\
\tablefoottext{*}{Frequency of the most intense transition of the hyperfine structure.}\\
\end{table*}

\subsection{IRAM and \herschel observations}
We observed several low-lying rotational transitions of \CCH and \CthHtw at  mm wavelengths using  
 the \IRAMthm telescope. We completed this dataset with \herschel \citep{pilbratt10} observations of a higher excitation line of \CCH and  two lines of \CH.  A summary of the spectroscopic parameters for all the observed lines are given in Table \ref{tab:spec}.
Throughout the paper,  we will use main beam temperature as intensity scale.
 Offset positions are given relative to the ionization 
front (IF: $\alpha_{J2000}$=06h07m46.2s, $\delta_{J2000}=-06^\circ23'08.3''$). 
For all the observations, the OFF position was chosen  to be free of molecular emission ([+10\arcmin,0\arcmin] for \herschel; [+400\arcsec, -400\arcsec] for IRAM).
A summary of the observations is shown in  Table \ref{table_obssum}.

The 3mm and 2mm observations were performed in July 2012 at the \IRAMthm telescope in Pico Veleta (Spain). 
We observed  the \CCH multiplet at 87\GHz and the \CthHtw line at 85\GHz using the EMIR receivers with the VESPA correlators, covering  a 180\arcsec$\times$180\arcsec\, region centered toward the IF. 
The 1~mm observations were performed in two different sessions using the HERA 3x3 dual-polarisation  receiver and in a successive run with the EMIR receivers.   During the first run (HERA) we obtained a 150\arcsec$\times$150\arcsec\ map of the \CCH multiplet at 262\GHz and the \CthHtw line at 217\GHz. 
The second observing run (January 2012) was part of a 2-D spectral survey at 1mm \citep[][in prep.]{trevino12}. The resulting \CthHtw maps  have a lower signal-to-noise ratio (SNR) and the field of view covered  in this second run was smaller, 2\arcmin$\times$2\arcmin.

The \herschel observations were obtained with 
the Heterodyne Instrument for the Far Infrared \citep[HIFI, ][]{degraauw10} 
in the context of the WADI guaranteed-time key program \citep{ossenkopf11}.  
The CH lines were observed with both the wide band spectrometer (WBS) 
and the high-resolution spectrometer (HRS). They provide spectral  resolutions (at 500\GHz) 
of $\Delta {\rm v}$ = 1.2\kms  and 0.3\kms, respectively,  allowing to resolve the line profiles. Cross-calibration between the two backends and between the two polarizations indicates an uncertainty of $\sim 20\%$.
The CH (J = 3/2-1/2) and CCH (N=6-5) observations were performed in on-the-fly (OTF) observing mode along the strip indicated in Fig.\,\ref{fig_mapmon2}  while the excited ($J=5/2-3/2$) line of CH was observed in single pointing observing mode towards the IF, with a double beam switching pattern. In this paper we also compare  the new observations with the [\Cp] strip already presented by \citet{pilleri12a}.

\subsection{Data reduction}

The basic data reduction of \herschel observations was performed using the
standard pipeline provided with the 
version 7.0 of HIPE\footnote{HIPE is a joint development by the Herschel 
Science Ground Segment Consortium, consisting of ESA, the NASA Herschel 
Science Center, and the HIFI, PACS and SPIRE consortia} \citep{ott10} 
and  Level 2 data were exported to the  FITS  format.
The \herschel and \IRAMthm{} data were  then processed  using the \GILDAS{}/\CLASS{} software\footnote{See
  \texttt{http://www.iram.fr/IRAMFR/GILDAS} for more information about the
  \GILDAS{} softwares.}  suite~\citep{pety05}.
 The spectra
were first calibrated to the \Tas{} scale using the chopper wheel
method \citep{penzias73}, and finally converted to  the main beam temperatures
(\Tmb) using the nominal forward (\Feff) and main beam (\Beff) efficiencies (in Table~\ref{table_obssum}, we show the value $\eta_l = {\Beff/\Feff}$).  The HIFI efficiencies have been recently calculated by \citet{roelfsema12}. The resulting amplitude accuracy is
$\sim$ 10\%  for both HIFI and IRAM observations. 

Visual inspection of the IRAM data revealed for all the observed lines 1)
the presence of more or less pronounced spikes at one quarter and three
quarter of the correlator window, 2) the presence of platforming in the
middle of the correlator window, 3) but otherwise clean baselines. Windows
around the  well-defined frequencies displaying spikes were set before
baselining. The line windows were defined on the spectra averaged over the
map, from [-1,24.5\kms] for the brightest lines to [6,16\kms] for the
faintest ones. To correct for the platforming, a zero-order baseline was
subtracted from each correlator sub-bands of each spectrum.  A
first-order baseline was needed for  \CthHtw. The bad channels, which occurred in a line-free section of the HERA spectra, 
were replaced by Gaussian noise
of same rms.  
The resulting spectra were  resampled and convolved with a Gaussian  to a map resolution larger than the HPBW of the telescope.  The map resolution is reported in Table \ref{table_obssum}. The spectral cubes were finally blanked where the
spectra weights (resulting from the gridding) were too low, giving the
fringed aspect of the map edges.

\section{Results}
\label{sec_results}

In this section, we describe the spectral and spatial variations of the  various molecular tracers. First, we present all the observations obtained towards the IF, which is one of the best-known position \citep[see, for instance,][]{fuente10, ginard12, pilleri12a}. We then describe the spatial/spectral variations along the \herschel strip (see Fig.\,\ref{fig_mapmon2}) and finally, we present the full maps.

\subsection{The ionization front}

\begin{figure*}
\centering
\includegraphics[width=0.85\linewidth]{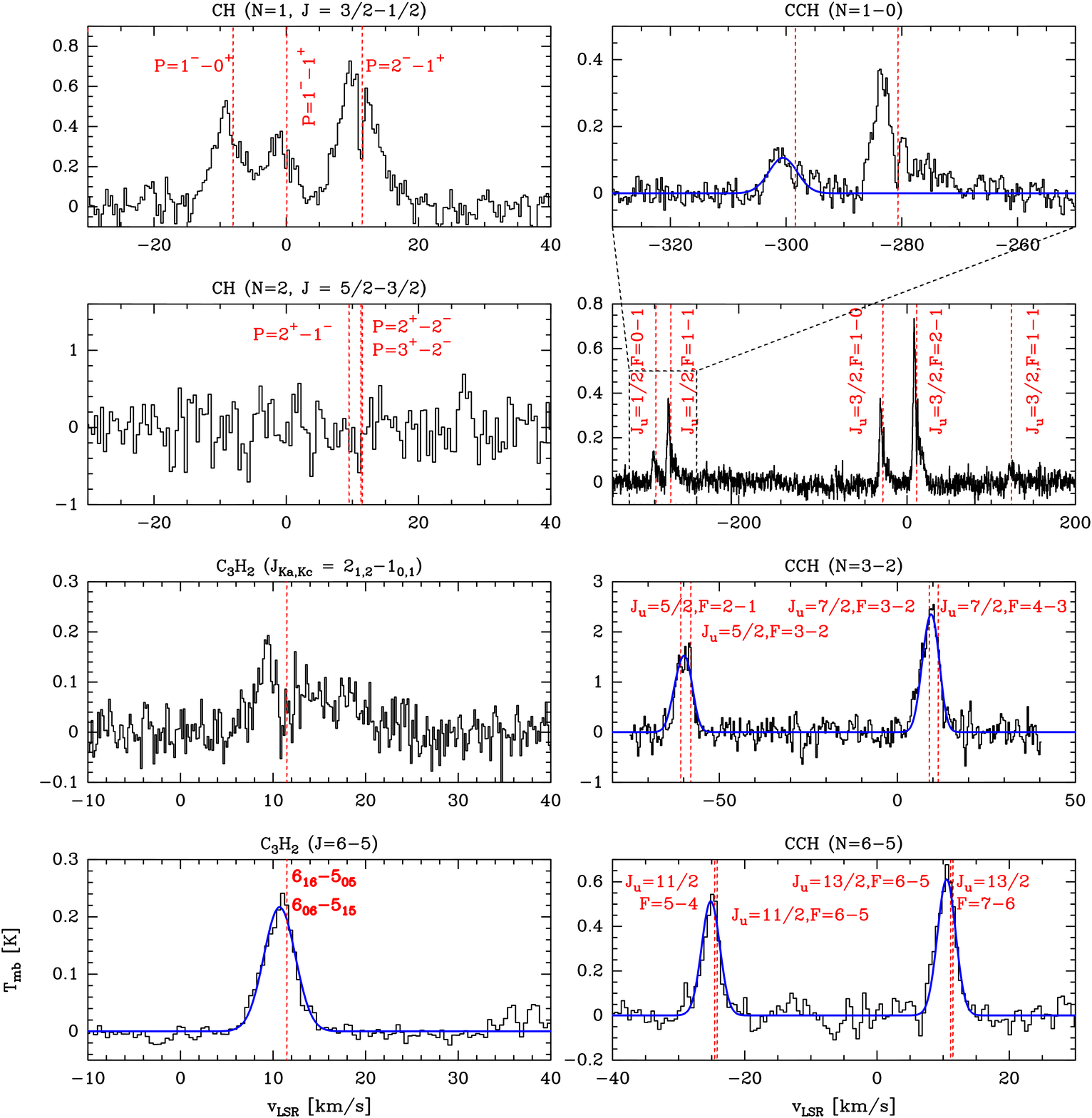}
\caption{Herschel and IRAM observations of \CCH and \CthHtw towards the IF. 1mm observations have been convolved to the spatial resolution of 39\arcsec. Solid blue lines show the  single-component Gaussian fits.  The velocity scale in each panel is relative to the most intense line in the hyperfine groups (see also Tables \ref{table_obssum} and \ref{tab:spec}).} 
\label{fig_obsiram}
\end{figure*}

In Fig. \ref{fig_obsiram}  we show the spectra of several \CH, \CCH and \CthHtw transitions  observed  towards the IF  with the \IRAMthm and \herschel. 
The lower-lying transitions show  a self-absorption feature at the ambient velocity of the cloud ($\sim 11$\kms). 
It is clearly detected  in the main hyperfine components of  CH, \CCH and \CthHtw, and coincides with the self-absorption detected in the [\Cp] strip observed with \herschel \citep{pilleri12a}. 
The self-absorption is barely visible in the faintest hyperfine components of the \CCH 1-0 line and is not detected in any of the higher energy transitions. The structure of the \CCH 3-2 line is due to the superposition of two hyperfine transitions with very similar frequencies that are not self-absorbed\footnote{This is suggested by the excellent quality of the fit of  the observed lines using a Gaussian function for each of the hyperfine component, and assuming a standard intensity ratio and frequency shift, see Table \ref{tab:spec}. }.
The lower-lying transitions show some emission in the red and blue wings, which may trace either an expanding PDR \citep{pilleri12a} or the molecular outflow \citep{tafalla97}. 

\subsection{The NE-SW strip}
\CH and \CeiO (5-4) have been observed only towards the NE-SW strip shown in Fig.\,\ref{fig_mapmon2}.
 This strip, already studied in \citet{pilleri12a}, passes through the IF and extend up to 1.5\arcmin\ in each direction, crossing the main PDRs of the region and extending to offsets where the emission is  dominated by the molecular envelope. However, the large beam of  \herschel at this frequency ($\sim 40$\arcsec)  results in even the positions [$\pm40\arcsec, \pm40\arcsec$] picking-up some emission from the innermost PDR. 

\begin{figure*}
\centering
\includegraphics[width=0.9\linewidth]{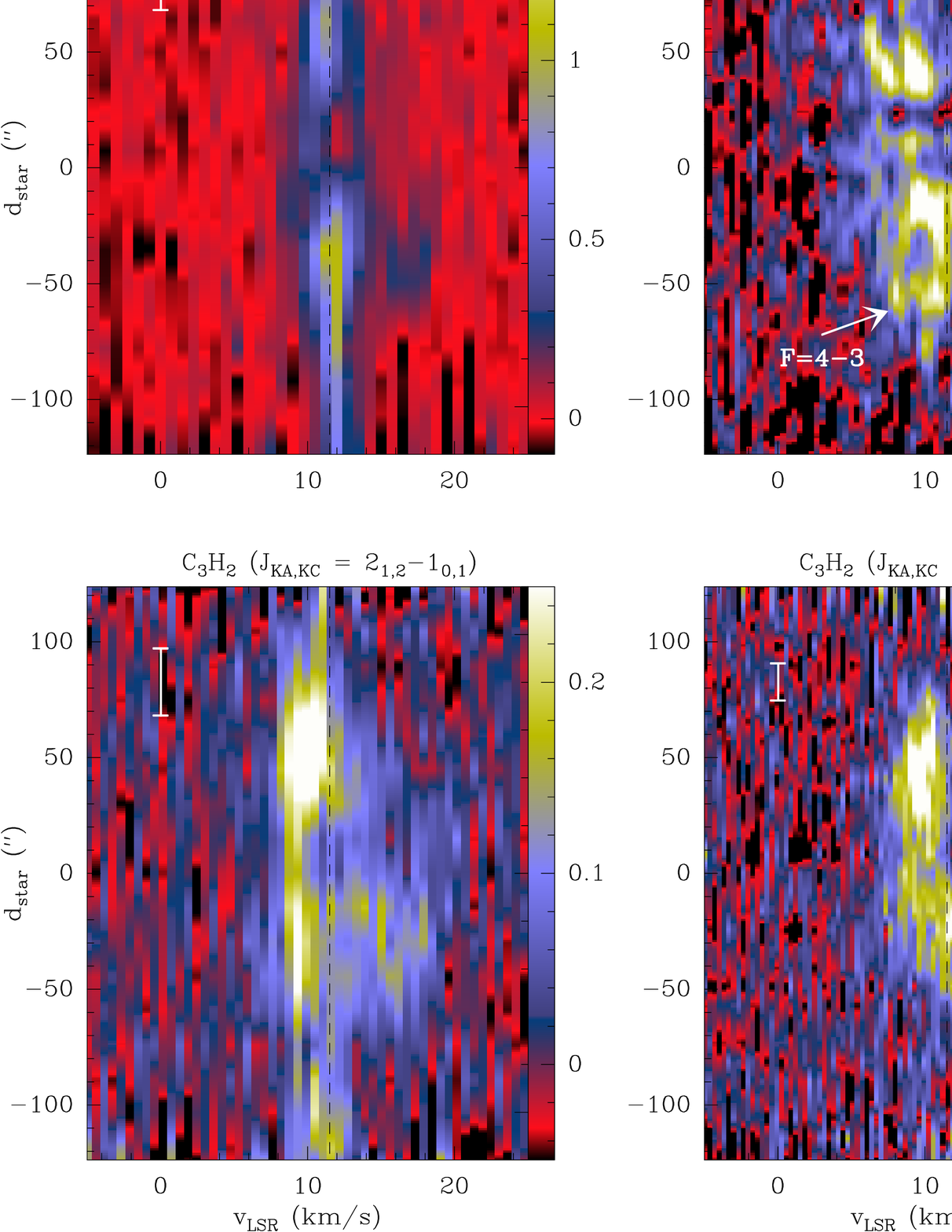}
\caption{PV diagrams of the \herschel and IRAM observations along the SW-NE strip. The \Cp diagrams comes from \citet{pilleri12a}. The double peaked structure of the \CCH lines is due to two hyperfine component of the same line, while the strong dip in both CH and [\Cp] are  partially due to self-absorption. The white line in the top-right corner of each diagram represents the HPBW of the observation. The black dotted line indicates $v_{LSR} = 11.5\kms$.}
\label{fig_pvs}
\end{figure*}

In Fig. \ref{fig_pvs} we show the position-velocity (PV) diagrams along the cut observed with \herschel, i.e. in the  direction perpendicular   to the molecular outflow  \citep{tafalla94, tafalla97}, which extends in the northwest-southeast direction from the IF. For  a comparison, we also display the PV diagram of the  [\Cp] line along the same cut \citep{pilleri12a}. For each transition, we show the diagram of  the component with the highest signal-to-noise ratio (SNR). Although  the data have very different  angular resolutions, the spatial variation of CH is in some way similar to that of [\Cp]: for instance, in both tracers the emission is extended up to 60\arcsec\, from the IF and the   main component shows  self-absorption through the whole cut. Both lines present a  secondary peak at a distance of $\sim10\arcsec$ at about 8\kms.  The line widths are also similar indicating that the CH emission, like [\Cp],  originates in the  innermost layers of the expanding central PDR. 

 The PV diagram of  \CCH 1-0 is very similar to the \CeiO 2-1 line in terms of the spatial extension and line width. They are both detected  along the entire strip, and their typical line width is about 2.5\kms. However, the  \CCH line toward the IF  displays self-absorption at central velocities and emission in the red wing. 
Similarly to the \CCH 1-0 line, the \CthHtw $2_{1,2}-1_{0,1}$ transition shows self-absorption at 11\kms  and emission in the red wing toward the IF. However, it is less spatially extended and therefore  likely  associated  with  the innermost layers of the PDR. 
 The higher-energy transitions of all species are  less spatially extended compared to [\Cp] and \CH and do not show self-absorption. 
Finally, there is a systematic shift of the peak velocity of about 2\kms going from the south-west towards the north-east of the IF, which has been attributed to either an asymmetry in the expansion pattern or to a large-scale slow rotation of the entire cloud \citep{loren77, pilleri12a}.

\subsection{Maps}

\begin{figure*}
\centering
\includegraphics[width=0.49\linewidth]{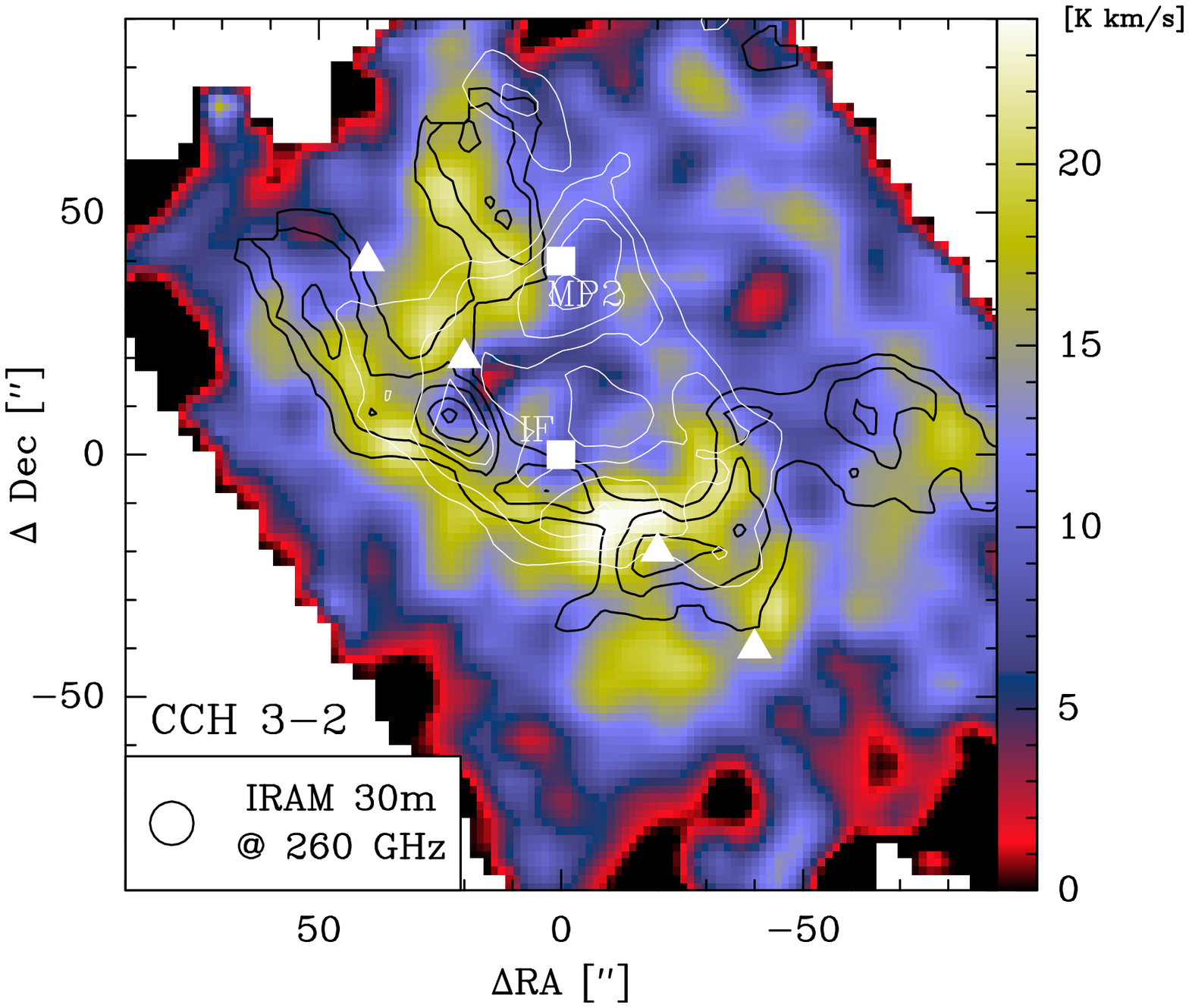}\hfill
\includegraphics[width=0.49\linewidth]{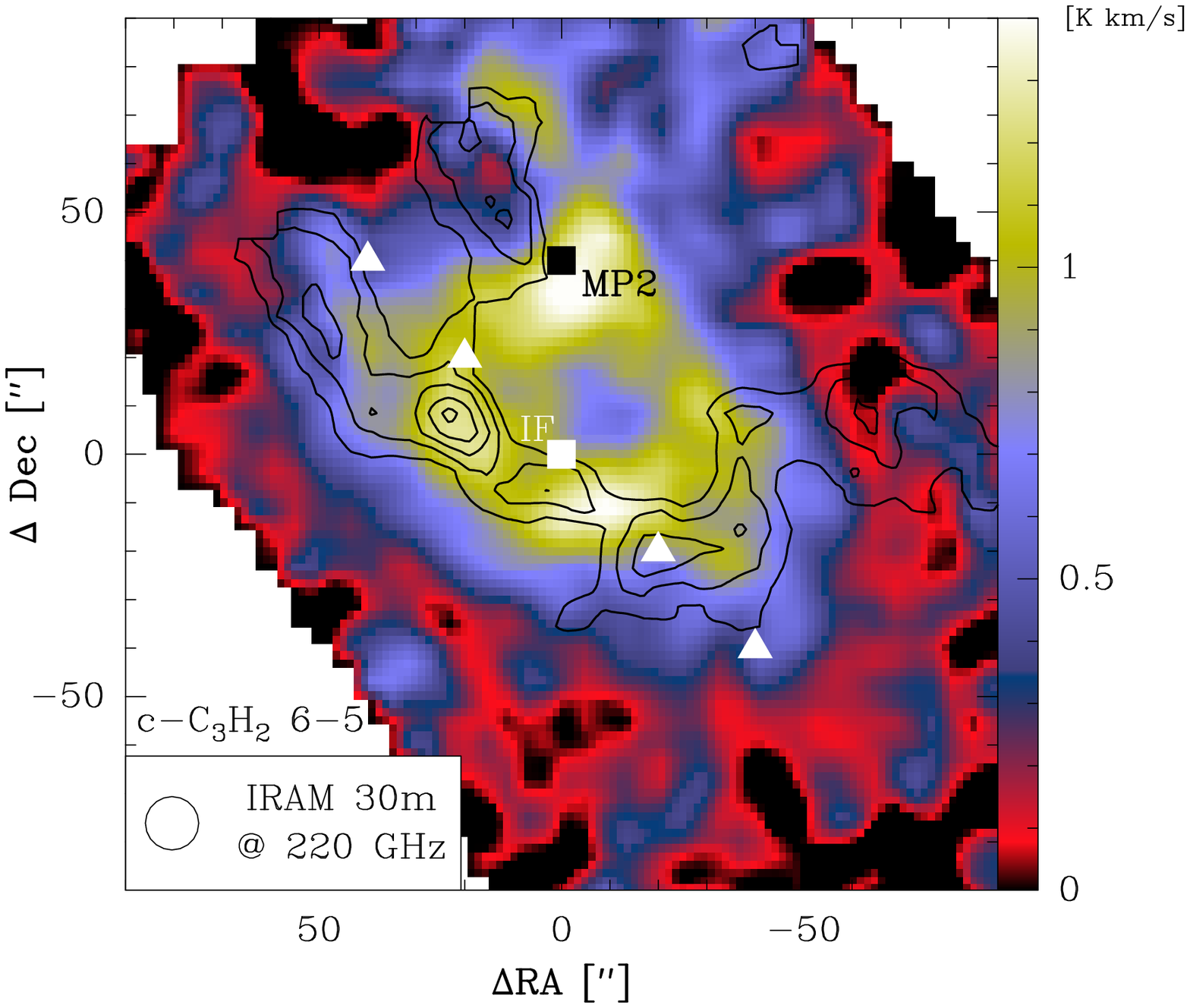}
\caption{Maps of integrated intensity (in the range 5-15 \kms) for \CCH (262.004\GHz) and \CthHtw (217.822\GHz) obtained with the EMIR  receivers. Black contours are the \CeiO line shown in Fig.~\ref{fig_mapmon2}. White triangles and squares represent the single positions studied in this paper.  White contours in the left panel represent the \CthHtw line integrated intensities (in the range 5-15\kms) starting from 0.6\kms and in linear steps of 0.2\kms.}
\label{fig_mapshydroc}
\end{figure*}
The maps carried out with the 30m telescope allow us to investigate about the spatial distribution of \CCH and \CthHtw.
Figure \ref{fig_mapshydroc} shows the spatial distribution of the integrated line intensity  of the \CCH 3-2 and \CthHtw 6-5 lines between 5 and 15\kms  observed with HERA\footnote{The EMIR maps of all the observed lines are shown in Fig.~\ref{fig_images_convolved}.} at the spatial resolution of 10\arcsec.  
The spatial pattern of \CCH is very similar to that of  \CeiO 2-1 (Fig.\,\ref{fig_mapmon2}): the emission to the northeast and southwest  extends up to 1\arcmin\, from the IF, and  peaks at a distance of $\sim30\arcsec$. The emission is opened to the north and the northwest, following the cometary shape of the molecular bar.
The \CthHtw emission is more compact, with the center at an offset [-10\arcsec, +10\arcsec] relative to the IF. The emission extends less than 40\arcsec\, from the center, and peaks at a distance of about 10\arcsec\, south from the IF. The line intensity  also  displays a peak toward the MP2 position, which does not have a correspondence in the  \CCH map. The  emission to the north is  more extended compared to \CCH, following the IRAC 8\mum emission \citep{ginard12}.

\section{Column densities and abundances}
\label{subsec_coldens}

In this section, we  derive the column densities and abundances of \CH, \CCH and \CthHtw. First, we concentrate in the SW-NE strip, where we use the \herschel observations to study \CH and the excitation of \CeiO. We study in more details the excitation of \CCH using a two-phase LVG model towards the IF and study the emission in the line wings. Finally, we extrapolate these results to the analysis of the full maps to obtain more information on the spatial distribution of these tracers.  The self-absorption features are discussed in Appendix \ref{appen_abs}.

\subsection{The SW-NE strip}

\begin{table*}
\begin{center}
\footnotesize
\caption{Integrated intensity (using \Tmb scale) between 5 and 15 \kms, after smoothing the data to the larger $HPBW$ (40\arcsec\, for \CCH and \CeiO , 28\arcsec\, for \CthHtw).  Non-detections are given at a 3$\sigma$ level assuming a line width of 1\kms.
See Table \ref{tab:spec} for the spectroscopic parameters.} 
\label{obsinteg}
\begin{tabular}{l|c|cc|cc|c}
\toprule
				& IF	[0\arcsec; 0\arcsec]				&  [20\arcsec; 20\arcsec]&  [40\arcsec; 40\arcsec]&    [-20\arcsec; -20\arcsec]&  [-40\arcsec; -40\arcsec] &  MP2 [0\arcsec; 40\arcsec] \\
\midrule
Transition 	&  Area &  Area &  Area &   Area &  Area & Area \\
		         & [\K\kms] 	             		 		         & [\K\kms] 	             		& [\K\kms] 		         & [\K\kms] 	 & [\K\kms] 	 & [\K\kms] 	             	     	\\
\midrule
\CeiO 2-1		& 14.56 						& 15.55		& 15.71		& 15.68	& 13.12	& 13.90 \\
\CeiO 5-4		& 13.04 						& 13.38		& 13.26		& 12.12	& 11.03 & -  \\
\midrule
\CH$_{J=3/2-1/2, P =1^--0^+}$ [5-12]\tablefootmark{(a)}		&	1.81		& 1.43		& 0.62		&1.03	& 0.22 & - \\
\CH$_{J=3/2-1/2, P =2^--1^+}$ [12-15]\tablefootmark{(a)}		&  1.19		& 0.74		& 0.36		& 0.95 	& 0.22 & - \\
\CH$_{J=5/2-3/2}$							&$\lesssim0.84$ & - & - & - & - & -\\
\midrule
\CCH$_{N = 1-0,\,F=0-1}$						&  0.98		& 1.35		& 1.39		& 1.61	& 1.87	& 1.29\\
\CCH$_{N = 3-2,\,J_u = 5/2, F=4-3\,\&\,F=3-2}$ 	& 13.75	& 13.85	&11.70		&15.63	&11.99	 & 	12.19\\
\CCH$_{N= 6-5,\,J_u = 13/2, F=7-6\,\&\,F=6-5}$ 	& 4.97				& 4.97		&4.05		&5.02 	& 3.73	 & - 	\\
\midrule
\CthHtw$_{212}$	& 0.73					&	1.55		& 1.71		&1.73	& 1.32 & 1.71	\\
\CthHtw$_{404}$ 	& 0.45					& 0.44		& 0.44		& 0.45	&0.19	& 0.67	\\		
\CthHtw$_{414}$	& 1.42					&  1.71		& 1.34		&1.52	 &0.80 & 2.0 	\\
\CthHtw$_{432}$	&  0.63					& 0.46		& 0.37		& 0.49	&0.32	& 0.53	\\
\CthHtw$_{514}$	& 0.76					& 0.56		& 0.33	 	& 0.46	& 0.20       & 0.87	\\
\CthHtw$_{616\,\&\,606}$	& 0.99					& 1.00	&0.52		&0.85 	&0.42	& 1.01\\
\CthHtw$_{625}$	& 0.57					&  $<0.3$		& $<0.3$		& $<0.3$	& $<0.3$	& 0.41\\
\CthHtw$_{707\,\&\,717}$	& 0.69					& 0.61		&$<0.4$		&0.58	& $<0.3$	&0.78  \\
\bottomrule
\end{tabular}
\end{center}
\tablefoottext{a}{The numbers in brackets indicate the velocity range over which the line has been integrated to avoid self-absorption or contamination from a satellite line.}
\end{table*}

\newcommand{\delvc}{\ensuremath{\Delta v_{c}}\xspace}
\newcommand{\delvr}{\ensuremath{\Delta v_{r}}\xspace}
\newcommand{\delvb}{\ensuremath{\Delta v_{b}}\xspace}

\subsubsection*{\CeiO}
To obtain an estimate of the excitation temperature and column density of \CeiO along the \herschel strip, 
we have combined our J=2-1 observation with the J = 5-4 observations published in \citet{pilleri12a} and with the J = 1-0 observations of \citet{ginard12}. We have convolved the data to the  lowest spatial resolution (40\arcsec\, for the J = 5-4)  and built rotational diagrams for the five positions along the strip. The J=1-0  data cannot be convolved, since  they were single pointing observations, but the spatial resolutions are not very different (29\arcsec\, vs 40\arcsec) and the \CeiO emission is very extended, so  that the emission fills the beam at both spatial resolutions. The corresponding rotational diagrams are shown in Fig.\ref{fig_rotdiacch}. 
 The results are consistent with those of \citet{ginard12}: the rotation temperature slightly decreases from $\sim 37$\K toward the IF to $\sim 31$\K at a distance of 50\arcsec. We obtain a column density of \ex{8.3}{15}\cmq toward the IF, and slightly (20\%) higher values at larger offsets. These values are consistent within a factor of 2 with the geometrical model presented by \citet{pilleri12a} to fit the high-J lines of CO, \thCO and \CeiO. This model consists of a spherical cloud with  an $\Av=8$  PDR and  an $\Av=40$ low-density envelope.  This geometrical model yields a factor of 2 higher column densities compared to the \CeiO values, assuming an isotopic ratio $^{16}$O/$^{18}$O = 500 and a standard abundance of CO relative to \HH of $ \ex{1}{-4}$. In the following, we assume therefore an uncertainty of a factor of 2 in the column density of \HH and  in the abundances of the hydrocarbons. However, the absolute value of the column densities, and consequently the relative
abundances, are not affected by this uncertainty.

\subsubsection*{\CH}
As shown in Fig.\,\ref{fig_obsiram}, the  main component of CH is  self-absorbed at the ambient cloud velocity and  the line wings of the different components overlap.
 To obtain the integrated intensity of the \CH line, we used the $P=1^--0^+$ and the $P=2^--1^+$ transitions for the velocity intervals  [5-12]\kms and [12-15]\kms, respectively. 
Using the first satellite to obtain the area in the blue interval we avoid the self-absorption that affects the main component. On the other hand, the red part of the $P=1^--0^+$ line is contaminated by the overlap with the second satellite, so that for this interval we have to rely on the main component. The integrated areas for each interval are shown in Table \ref{obsinteg}.

 When computing the \CH column density,  we face the difficulty that there are no collisional rates available for CH, so that  one can only assume a given rotational temperature \Trot per velocity interval. 
The non-detection of the CH triplet at 1656 GHz provides an upper limit to \Trot if we assume LTE conditions, so that both transitions share the same rotation temperature. This gives upper limits for \Trot of  25\K (blue)  and 30\K (red) towards the IF. 
Considering the high dipole moment of CH and the physical conditions in Mon~R2, we can estimate a lower limit to the rotational temperature to be about 7\K.
 Assuming that the emission is optically thin and that  it fills the beam,  we derive a column density of $\ex{1.2^{+1.9}_{-0.6}}{14}$\cmq towards the IF for $\Trot = 10$\K, with the uncertainties driven by upper and lower limit in \Trot. At the  [-40\arcsec; -40\arcsec] position,  the \CH column density is a factor of 10 lower, $\ex{1.7^{+3.1}_{-1.1}}{13}$\cmq, assuming  $\Trot = 10$\K. At this position, \Trot is likely lower compared to the IF, due to a decrease in kinetic temperature and local density, which yields an overestimate of the column density by a factor of 2 for $\Trot = 7$\K. At  offsets  larger than 60\arcsec, the ground state transition of \CH is not detected, implying that the column density is even lower. Thus, we consider the column density at [-40\arcsec; -40\arcsec]  as an upper limit of the \CH abundance in the envelope. 

Using the H column density derived through \CeiO ($N({\rm H}) = \ex{9}{22}$\cmq), we estimate a \CH
abundance relative to H nuclei of $\sim \ex{1.5}{-9}$ toward the IF   decreasing to $\sim \ex{3}{-10}$ at an offset of [-40\arcsec, -40\arcsec], for $\Trot = 10$\K. 
This is consistent with most of the CH emission at central offsets  arising from the central expanding PDR.
 However, if the bulk of  \CH emission arises indeed in the PDR layers, using \CeiO as a probe of \HH most likely underestimates the abundance because $i)$ the abundance of \CeiO is affected by photodissociation and $ii)$ because \CeiO may not coexist with the CH present in the gas where \CeiO is partly photo-dissociated. If we consider that the emission stems only from the innermost 10 mag in each side of the spherical cloud \citep[$N({\rm H}) = \ex{4}{22}$\cmq, cf. Sect. \ref{subsec_madex}, \ref{chemicalmodel} and][]{pilleri12a}, the mean abundance in the PDR would be 
of $\ex{3.3^{+6.0}_{-2.2}}{-9}$  relative to H nuclei, with the error bars  driven by the uncertainty in \Trot.

\subsubsection*{\CCH}
The \CCH line at 87\GHz is composed of  6 hyperfine components. The most intense line is self-absorbed at a velocity of $\sim 11$\kms. To estimate the molecular column density, we used the faintest  observed component (87.407\GHz), which is less affected by self-absorption. 
We have observed four hyperfine components of the  J = 3-2 and 6-5 multiplets, which are blended in pairs of two. For each  group, we derived the total integrated intensity of the lines and then scaled it according to the relative strengths  to estimate the contribution of each blended lines separately. 

Figure \ref{fig_rotdiacch} shows the rotational diagrams at several position along the \herschel cut. We have convolved the J = 3-2 line to the spatial resolution of the 6-5 observations (39\arcsec).  For all the lines, we assumed a 30\%   uncertainty for the intensity, except for the 1-0 line where we have assumed an  uncertainty of a factor of 2 due to a possible self-absorption. For the 3mm transitions we used the hyperfine fit task 
inside CLASS, and found that the relative line intensities of the 3mm transitions are consistent with optically thin emission. For the higher frequencies, the hyperfine lines overlap and this method is not reliable. However, our LVG results (see below) show that the opacities of these lines are also lower than 0.1.  

One single component cannot fit the rotational diagram at any of the positions. For this reason, we  used a two-components rotational diagrams to fit the integrated intensities of all the lines along the strip.  The warm component has a rotational temperature of $\sim 35$\K while the cold component is at about 10\K.  These temperatures are consistent with the superposition of two phases  along the line of sight: a dense and high-UV illuminated PDR  and a low-UV irradiated envelope \citep{pilleri12a}. 

The \CCH column density is approximately constant  ($\sim \ex{7.5}{14}$\cmq) along the entire strip, but the column density of the hot component decreases by about 25\% at  larger offsets, where the emission is dominated by the envelope.
This supports the interpretation of the warm component being associated with the PDR around the \HII region (hereafter, we will refer to this component as the PDR component).
 It is important to note that this gradient is a lower limit to the real one because  these column densities have been determined at a low spatial resolution (40\arcsec), thus diluting higher gradients in their distribution that may be present in the beam.
Using the \HH column density derived  from \CeiO, we estimate a mean \CCH abundance of \ex{9}{-9} relative to H nuclei.%

\subsubsection*{\CthHtw}

The \CthHtw   2$_{1,2}-1_{0,1}$ transition is self-absorbed towards the IF implying that we can only establish a lower limit to the column density of \CthHtw at this position. Elsewhere, the spectra are not self-absorbed and we can derive accurate column densities.

We convolved the higher-J transition to the spatial resolution of the  2$_{1,2}-1_{0,1}$ line, and used the rotational diagram technique to  compute the corresponding column densities. The rotational diagrams show that the column density decreases at larger offsets, from $\sim \ex{6.0}{12}$\cmq towards the IF to $\sim \ex{4.3}{12}$\cmq at an offset of [-40\arcsec,-40\arcsec] to the SW.  Only one component is needed to fit reasonably well the observed intensities, with a rotational temperature in the range 8-16\K. 
The excitation temperature seems to decrease farther from the IF. 
 The spatial distribution of this species suggest that its emission is dominated by the dense PDR around the \HII region, similarly to
the warm component of CCH. The difference between the rotation temperature of  \CthHtw and that of the hot component of CCH is partially due to the higher different dipole moment of \CthHtw, which makes it much more difficult to excite by collisions. This is also consistent with the results of the LVG modeling in Sect. \ref{subsec_madex}.  

 Again, using \CeiO as a tracer of the \HH column density yields an abundance relative to H of $X(\CthHtw) =  \ex{7.3}{-11}$  towards the IF, decreasing by about 30\% at the position $[-40\arcsec,-40\arcsec]$.  
As for \CCH,  more details can be obtained assuming a rotational temperature and using only the 1mm observations, improving the spatial resolution by a factor of 2.5. Such improvement will be shown in Sect.\,\ref{subsec_spatialdist}.

\subsection{Two-phase decomposition towards the IF}
\label{subsec_madex}
The \CCH rotational diagrams indicate that there are two emission components  towards the IF.  This is also consistent with the source model proposed by \citet{pilleri12a}, in which a very dense PDR ($\nhh = \ex{3}{6}$\cmc) is surrounded by a lower density envelope ($\Tkin = 35$\K, $\nhh = \ex{5}{4}\cmc$). In the previous section we interpreted the warm component of \CCH and the emission of \CthHtw as arising in the PDR, with the cold component arising from the envelope. 
 In order to verify this assumption and  quantify the excitation effects on \CCH and \CthHtw, we used the MADEX large velocity gradient (LVG) code \citep{cernicharo12} to reproduce the  \CCH and \CthHtw emission towards the IF using the collisional coefficients of \citet{spielfiedel12} and \citet{chandra00}, respectively.
 
We only have  a small number of transitions with different upper energy levels, and therefore we cannot  accurately constrain all the physical parameters (\nhh, \Tkin, $N$). Therefore, 
we have to rely on previous knowledge of the source structure \citep{pilleri12a} to fix the  temperature and density in each of these two phases and fit the column densities. We assumed densities of \nhh$=$\ex{3}{6}\cmc and \ex{5}{4}\cmc and kinetic temperatures of 70\K and 35\K for the PDR and the envelope, respectively.  The line width for both components has been set to 2\kms. By assuming these parameters and varying the column densities in each of the two phases, we obtain the best fit to the observations with the solutions shown in Table \ref{tab_lvg}.  The line opacities returned by MADEX are lower than 0.1 for all the transitions. For \CCH, we reproduce all the observed intensities within a 10\%  uncertainty. For \CthHtw, we are able to reproduce within a 15\% error the observations with the best  SNR, while the other transitions are fitted within 30\%. 
This approach gives 
the column densities and abundances reported in Table \ref{tab_lvg}.
 The results of MADEX modeling for all the transitions observed toward the IF are shown on the rotational diagrams in Fig.\,\ref{fig_rotdiacch}. The uncertainties in the abundances are dominated again by the uncertainty in the total H column density, and can be estimated to be a factor of 2.

\begin{table}[htdp]
\caption{Results of the two-phase LVG modeling towards the IF. Abundances are relative to H nuclei and are calculated assuming that the PDR  has a thickness of $\Av =8$ and the envelope  of $\Av = 40$ on each side of the spherical cloud. } 
\begin{center}
\begin{tabular}{l|cccc}
\toprule
				& 		Combined		&	PDR			&	Envelope		\\
\midrule
$N(\CCH)$ [\cmq]	&		\ex{1.0}{15}	&	\ex{5.0}{14}		&	\ex{5.0}{14}	\\
$X(\CCH)$		&		\ex{5.7}{-9}	&	\ex{1.7}{-8}	&	\ex{3.5}{-9}\\
$N(\CthHtw)$ [\cmq]	&		\ex{6.0}{12}	&	\ex{4.0}{12}		&	\ex{2.0}{12}	\\
$X(\CthHtw)$		&		\ex{3.5}{-11}	&	\ex{1.4}{-10}	&	\ex{1.4}{-11}	\\
\CCr				&		0.006		&	0.008	& 	0.004			\\
\bottomrule
\end{tabular}
\end{center}
\label{tab_lvg}
\end{table}%

 \subsection{Emission in the high-velocity wings}
The ground-state transitions of both \CCH and \CthHtw\  show a broad emission at red-shifted velocities, which is not clearly detected in the higher-energy lines (see Fig.\,\ref{fig_obsiram}). Toward the IF, the integrated intensity in the interval [15-25]\kms is 0.64\Kkms and 0.44\Kkms for \CCH and \CthHtw, respectively.  This allows to estimate an upper limit to the gas density of the  region emitting in the wings  from LVG models  that fit the intensity of the 3\,mm lines and stay below the detection limit for the other transitions. This yields an upper limit to the gas density of $\nhh = \ex{1}{5}$\cmc for $\Tkin = 70\K$. Assuming this density, we derive column densities of \ex{3.5}{13}\cmq and \ex{1.5}{12}\cmq for \CCH and \CthHtw, respectively. 

We repeated the rotational diagram for  the \CeiO lines  in this velocity interval to estimate the column density of the gas in this region, finding $N(\CeiO) = \ex{1.9}{14}$\cmq, which corresponds to $N(\rm{H}) = \ex{1.9}{21}$\cmq, or about 1 mag. This can be used to obtain an estimate of the abundance of \CCH and \CthHtw in the region emitting in the wings, which gives \ex{1.8}{-8} for \CCH and \ex{7.8}{-10} for \CthHtw. 
  If the density was lower (e.g. $\nhh = \ex{1}{4}$\cmc), the column density required to reproduce the wing intensity would be 15 \% higher  for \CCH and a factor of 4 higher for \CthHtw, yielding a higher \CCr ratio.

\subsection{Spatial distributions}

\begin{figure*}[ht]
\centering
\includegraphics[width=0.9\linewidth]{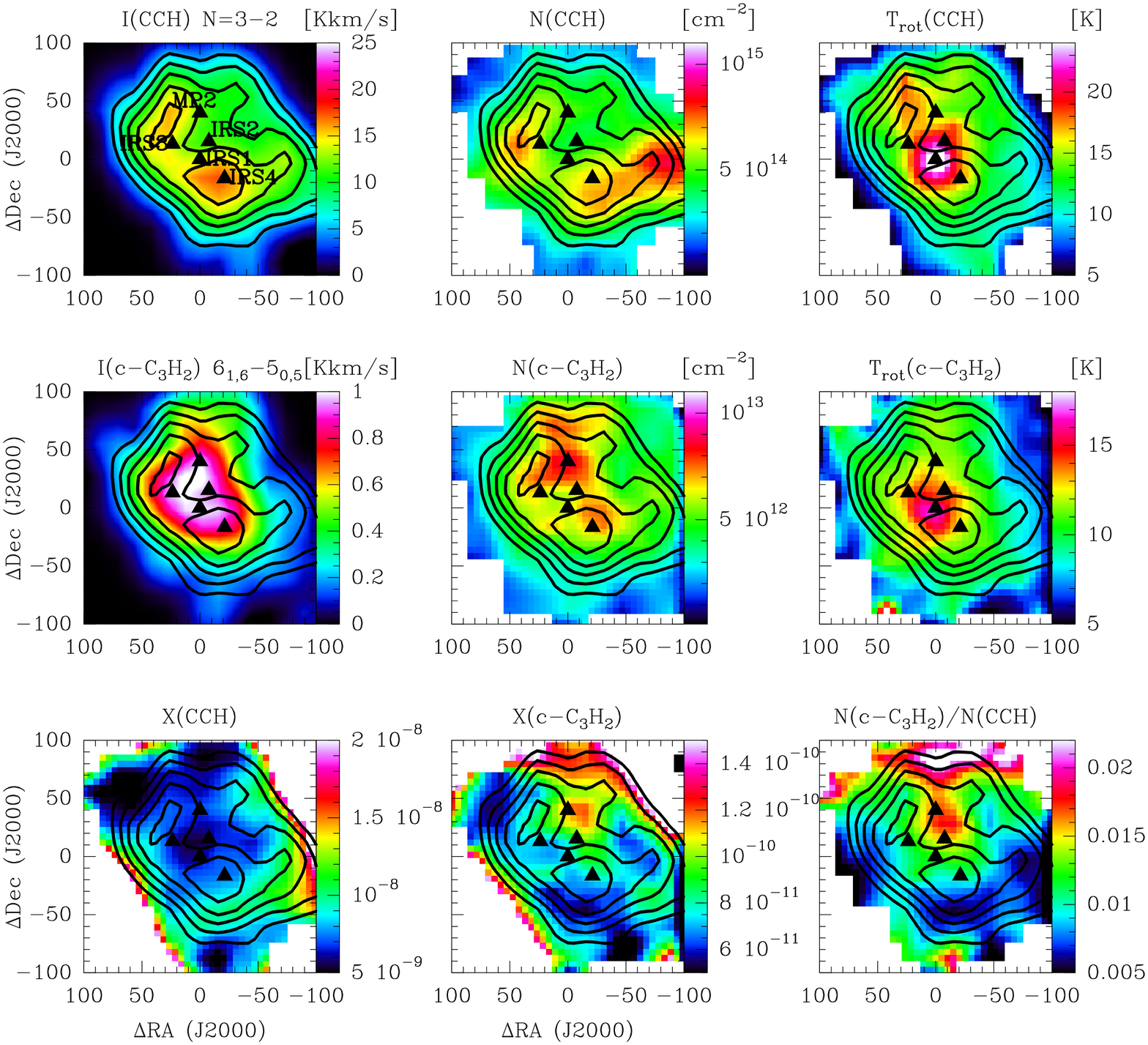}
\caption{{\it First two rows:} Maps of the integrated intensity of the \CCH\,  (262.004\GHz) and \CthHtw\,  (217.822\GHz) lines, and of the column density and rotation temperature derived applying the rotational diagram technique to each pixel.  Black contours indicate the \CCH 3-2 integrated intensity, and go from 5\K\kms to 20\K\kms in linear steps of 2.5\K\kms. The position of the infrared sources IRS1/2/3/4 \citep{henning92} and the MP2 position \citep{ginard12} are shown as black triangles. All the maps have been convolved to a spatial resolution of 29\arcsec. {\it Bottom row:} \CCH and \CthHtw abundance maps, obtained using \CeiO as a reference and  assuming $^{16}$O/$^{18}$O = 500 and a CO abundance of \ex{5}{-5} relative to H nuclei. The \CCr ratio is also shown. }
\label{fig_overplot_coldens}
\end{figure*}

\label{subsec_spatialdist}

In the previous paragraphs we studied the excitation of single positions for all the species and derived some patterns in both the column density and rotational temperature. Here, we generalize the study to the mapping observations. First, we convolved all the maps to the   lowest spatial resolution in the mapping dataset (29\arcsec). For each pixel  in the map, we constructed a rotational diagram for each molecule and derived an excitation temperature and a column density for both \CCH and \CthHtw. 
For simplicity, we used only one slope also for CCH. The results are shown in Fig. \ref{fig_overplot_coldens}. 

 For \CeiO, our analysis is limited by the fact that the J = 5-4 line was observed only in the NE-SW strip. We have therefore to assume a given \Trot in each pixel based on this dataset. 
We have interpolated the mean \Trot  obtained along the strip to obtain the rotation temperature of \CeiO as a function of the distance from the IF. Then we derived the corresponding column densities (Fig.\,\ref{fig_overplot_coldens}) assuming LTE.  Since the \CeiO  emission is optically thin, this yields an estimate of the \HH column density and allows to normalize the column densities obtained for \CCH and \CthHtw to derive their abundances. The corresponding maps are shown in Fig.\,\ref{fig_overplot_coldens}.

\begin{figure}[ht]
\centering
\includegraphics[width=1\linewidth]{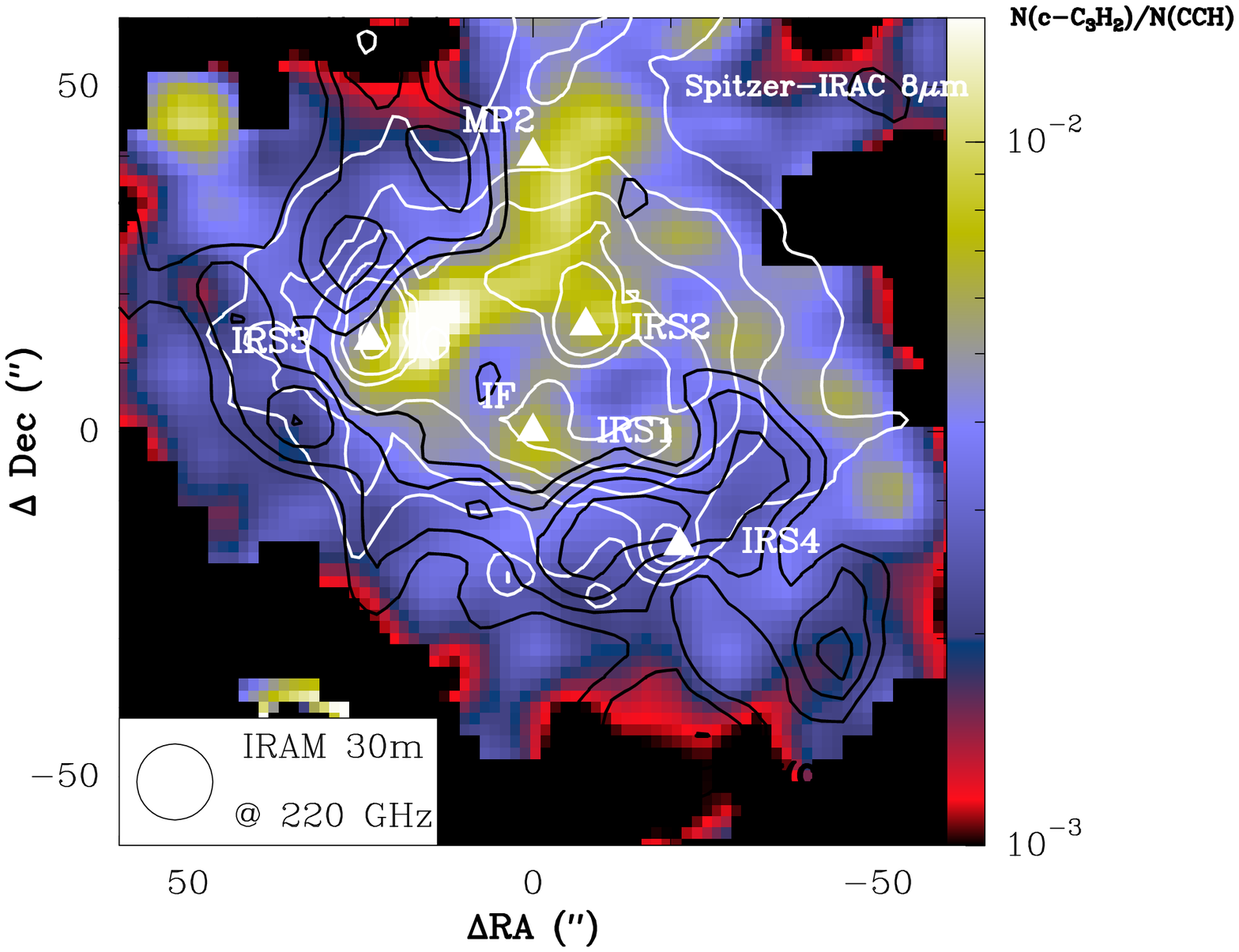}
\caption{Color image shows the ratio of \CthHtw to \CCH column densities  as derived from 1mm observations, assuming the rotational temperature of Fig.\,\ref{fig_overplot_coldens}. Pixels  without a detection of both \CCH and \CthHtw have been masked. The black contours represent the integrated intensity of the 262.004\GHz CCH line  (from 15 to 25 \Kkms in steps of 2.5\Kkms), and white contours represent the Spitzer IRAC 8 \mum emission.  }
\label{fig_overplot_irac}
\end{figure}

To have a deeper insight into the hydrocarbon chemistry we produced a map of \CCr, the ratio of the \CthHtw to \CCH column densities. 
The \CCr ratio shows variation up to a factor of 3  across the whole region, being minimum in the bulk of the molecular cloud, then increasing towards the PDR around the HII region and with the maximum located 40\arcsec\, towards the North, i.e.,  at the position of the low-UV PDR identified by \citet{ginard12}.
In fact, the \CCr seems to follow the emission of the hot dust (PAH and larger grains) traced by the 8$\mum$ emission. We can improve the angular resolution of our \CCr image by using directly the \CCH 3-2 and \CthHtw 6-5 images (both with an angular resolution of $\sim$10$"$)  to derive the \CCr but assuming the \CCH and \CthHtw rotation temperatures shown in Fig.\,\ref{fig_overplot_coldens} although these rotation temperatures were derived with a lower ($\sim$29$"$) angular resolution.  Note that the angular resolution of the IRAC 8\,$\mu$m image is 2\arcsec. This higher resolution \CCr image confirms the correlation with the 8$\mu$m image (see Fig. \ref{fig_overplot_irac}) corroborating the association of an enhanced \CCr with the external part of the PDR. In Sect.~\ref{subsec_linkpah} we will discuss the chemical implications of this correlation.

\section{Discussion}
\label{sec_discussion}

\subsection{Comparison with other PDRs}

 In the previous section, we have derived the abundances of small hydrocarbons in the different environments of Mon~R2. In the PDR, the abundances  are $X(\CH)=\ex{3.3}{-9}$, $X(\CCH)=\ex{1.7}{-8}$ and $X(\CthHtw)=\ex{1.4}{-10}$. 

Table \ref{tab_comparisonabund} shows the comparison of the \CCH and \CthHtw abundances in different PDRs. With the exception of the Orion Bar, the \CCH abundance is relatively constant in PDRs, in the range \ex{(0.6-1.4)}{-8}. The abundance in the PDR of Mon~R2 is consistent with these results. However, the envelope better matches  with the values  found in the Orion Bar \citep{fuente96, ungerechts97, fuente03}.
The case of the Orion Bar is interesting because, similarly to Mon~R2, it is a high-UV PDR with an associated envelope, the background molecular cloud not directly affected by the UV photons from the Trapezium cluster. The abundances for the Orion Bar shown in Table \ref{tab_comparisonabund} were estimated using low angular resolution observations (29\arcsec). The beam of these observations  comprises both the high-UV irradiated PDR and the warm envelope which
hinders from a reliable comparison with the values in other PDRs. Note, however,  that the \CCH abundance in the Orion Bar is similar  to the mean \CCH abundance in Mon~R2 (Table \ref{table_comparison}).

On the other side, the \CthHtw abundance shows a larger variation amongst the various PDRs. Whereas the PDRs with a relatively low \hab show an abundance of \ex{(0.6-1.1)}{-9}, NGC 7023 NW and Mon~R2 show much lower values, \ex{(0.4-2)}{-10}. This is also seen  as the trend in the \CCr ratio  in the different PDRs. Again, the values in the Orion Bar  are more similar to the mean value in Mon~R2. The \CCr ratio obtained in the PDRs of Mon~R2 is an order of magnitude lower compared to that found in "classical" PDRs \citep[see, for instance,][]{pety05} and in the diffuse medium \citep{gerin11}, and more similar to the value found in the dense gas surrounding the UC\HII region IRAS 20343+4129 \citep{fontani12}.

\begin{table*}[htdp]
\caption{Comparison of \CCH and \CthHtw abundances in different PDRs.}
\begin{center}
\begin{tabular}{lcccccc}
\toprule
			& 	\hab			&	\nhh	[\cmc]			& 	$X$(\CCH)\tablefootmark{(*)}			& 	$X$(\CthHtw)\tablefootmark{(*)}		&	\CCr 		& 	References	\\
\midrule	
 \multicolumn{7}{c}{MonR2}\\
\midrule	

MonR2 IF		&			&							&	\ex{5.7}{-9}	&	\ex{3.5}{-11}		&	0.006	&	a,b		\\
MonR2 MP2	&			&							&	\ex{2.6}{-9}\tablefootmark{(**,***)}	&	\ex{7.6}{-11} \tablefootmark{(**)}		&	0.029 \tablefootmark{(***)}		&	a,b		\\
MonR2  PDR	&	\ex{5}{5}	&		\ex{3}{6}					&	\ex{1.7}{-8}	&	\ex{1.4}{-10}		&	0.008		&		a		\\
MonR2 env	&	1-100	&		\ex{5}{4}				&	\ex{3.5}{-9}	&	\ex{1.4}{-11}		&	0.004	&		a		\\
MonR2 red wing	&		&						&	\ex{1.8}{-8}	&	\ex{7.8}{-10}		&	0.04		&		a		\\

\midrule	
 \multicolumn{7}{c}{Other PDRs}\\
\midrule	Horsehead	&	100			&	\ex{1}{5}							&	\ex{1.4}{-8}	&	\ex{1.1}{-9}		&	0.08		&	c				\\
Oph-W		&	400			&	\ex{2}{4}							&	\ex{6.0}{-9}	&	\ex{6.0}{-10}		&	0.1		&				d		\\
IC63			&	1500			&	\ex{1}{5}								&	\ex{1.1}{-8}	&	\ex{9.0}{-10}		&	0.08		&		d			\\
NGC7023		&	2600			&	$10^4-10^6$						&	\ex{6.0}{-9}	&	\ex{2.0}{-10}		&	0.03		&		e,f,g			\\
Orion Bar		&	\ex{2}{4}		&	\ex{1}{5}							&	\ex{2.0}{-9}		&	\ex{3.8}{-11}		&	0.01-0.03	&			f,h,i			\\
\bottomrule
\end{tabular}
\end{center}
\label{tab_comparisonabund}
\tablebib{(a): this work; (b): \citet{ginard12}; (c): \citet{pety05b} (d) \citet{teyssier04}; (e) \citet{fuente93};	(f): \citet{fuente03}; (g) \citet{pilleri12b};  (h): \citet{ungerechts97}; (i): \citet{fuente96}
}
\tablefoottext{{^*}}{ The uncertainty in the abundances is driven by the uncertainty in the \HH gas column density, which is assumed to be a factor of 2.}\\
\tablefoottext{^{**}}{ Calculated using the \CeiO observations of \citet{ginard12} as reference for the \HH gas column density: $N(\CeiO)=\ex{1.3}{16}$\cmq. }\\
\tablefoottext{^{***}}{ In MP2, we have to rely only on the 3mm and 1mm transitions of \CCH. Thus, we may underestimate the \CCH column density (and consequently, overestimate \CCr) by a factor of 2. }
\end{table*}%

\subsection{Comparison with chemical models}
\label{chemicalmodel}

In this section, we compare the observational results with the results of different chemical models,  in order to qualitatively distinguish which effects may be dominant in the different phases that co-exist in this region. To fully model the chemistry of this complex is beyond the scope of this paper.
Our goal is to explore the possible impact of different factors  such as the grain-surface chemistry or time dependency, on the carbon chemistry in these environments.

\newcommand{\Lpdr}{$L_{PDR}$\xspace}
\newcommand{\Lhd}{$L_{HD}$\xspace}
\newcommand{\Lenv}{$L_{env}$\xspace}

The PDR layers (hereafter, \Lpdr and \Lhd for the exposed PDR and the high density PDR, see Fig.\,\ref{fig_schema}) have different characteristics, in terms of physical conditions and time scales compared to the cold envelope. 
We assumed the density structure described in Sect.\,\ref{sec_observations} (see also the top panel in Fig.\,\ref{fig_chem}), in which the mainly atomic layer \Lpdr ($\Av<1$, $\nhh=\ex{2}{5}$\cmc, $\hab = \ex{5}{5}$) is followed by  the high density layer \Lhd  ($1<\Av<10$\,, $\nhh = \ex{3}{6}$\cmc), where many molecular species start to be abundant. These innermost layers are surrounded by the low-density envelope \Lenv ($10<\Av< 50$, $\nhh=\ex{5}{4}$\cmc). The assumption of constant density has to be taken with caution: local variations of gas density (clumps) and smoother gradients are expected to be found in the PDR layers, and the density in the envelope most likely decreases with the distance from the IF. However, this zero-order approximation allows to reduce the free parameters of the modeling to a few, which ease the comparison of models and observational data. 

To model the  \Lpdr and \Lhd  layers, we used the  Meudon PDR code  \citep{lepetit06,goicoechea07,gonzalez08,lebourlot12} version 1.4.3  using the density profile shown in Fig.\,\ref{fig_chem}. The Meudon Code computes the steady-state solution to the thermal balance and gas-phase chemical networks using accurate radiative transfer calculations and a plane-parallel geometry.  The transition between \Lhd and \Lpdr is calculated self-consistently by the code. This yields  stationary solution for the gas and dust temperature, as well as the abundance of each molecule in each of the layers by assuming equilibrium conditions. Time-dependent effects and gas-grain interactions are not included in the model and therefore its use is justified for the region where chemical equilibrium has already been reached and where gas-grain interaction is not expected to play a major role. However, the Meudon PDR code has very extensive gas-phase chemical networks and is best-suited for studying the gas-phase chemistry of Mon~R2.

To study the low-density, cold envelope we used the  UCL\_CHEM code \citep{viti04}. 
This code includes gas-grain interaction: for the purpose of this paper it is used as a two-phase model which consists  of the collapse of a prestellar core (Phase I), followed by the subsequent warming and evaporation of grain mantles due to the enhanced (with respect to the ISRF) radiation field (Phase II).
In phase I, a diffuse cloud of density 10$^2$ cm$^{-3}$ undergoes free-fall collapse until it has reached a final density (set by the user).
This phase occurs at a temperature of 10 K. During the collapse, atoms and molecules collide with, and freeze  onto, grain surfaces. The advantage of this approach is that the ice composition is not assumed, but it is derived by a time-dependent computation of the chemical evolution of the gas-dust interaction process which, in turns, depends on the density of the gas. Hydrogenation occurs rapidly on  grain surfaces, so that, for example, some percentage of carbon atoms accreting will rapidly become frozen out as methane, CH$_4$. 

 For all the models we use the same initial elemental abundances (see Table \ref{meudon}). The codes employ the reaction rate data from the UMIST astrochemical database,  augmented with grain-surface (hydrogenation) reactions as in \citet{viti04} and \citet{viti11} for the UCL\_CHEM  models. In both Phases of this code, non-thermal desorption is also considered as in \citet{roberts07}. 

\begin{table}[btph]
   \centering
      \caption{Input parameters for the Meudon PDR code}
   \begin{tabular}{lll} 
    \toprule
 	Parameter & & Value \\
	\midrule
	      $ G_0$    & Radiation field intensity &  \ex{5}{5} \\
	   $ G_0^{ext}$    & External Radiation field intensity & 100 \\
    $A_{\rm V}$& Total cloud depth &50 mag	\\
    Extinction & & Standard Galactic\\
    $R_V$	& $A_{\rm V}$/$E_{\rm B-V}$ & 3.1 \\
     $\zeta$	 & Cosmic ray ionization rate & \ex{5}{-17}~s$^{-1}$	\\
    $a_{min}$  & dust minimum radius& \ex{3}{-7} cm	\\
    $a_{max}$ & dust maximum radius & \ex{3}{-5} cm	\\
    $\alpha$ & MRN dust size distribution index   &3.5\\
    He/H & Helium abundance & 0.1  \\
    O/H & Oxygen abundance &  \ex{3.2}{-4}  \\
    C/H & Carbon abundance &  \ex{1.3}{-4}  \\
    N/H & Nitrogen abundance & \ex{7.5}{-5}  \\
    S/H & Sulfur abundance & \ex{1.8}{-5}  \\
     \bottomrule
   \end{tabular}
   \label{meudon}
\end{table}

\subsubsection{Abundances in the PDRs}
 The column densities and abundances in the PDR obtained from the observations refer to the innermost 10 mag of the PDR. Therefore, we have to compare them with the mean values in the \Lpdr$+$\Lhd layer.   
Another way to study the chemical difference between these species and minimize the uncertainties due to the assumed physical structure is to consider the column density ratios $N(\CH)/N(\CCH)$ and \CCr instead of absolute abundances. 

In the exposed PDR layer (\Lpdr), the gas temperature and UV field are so high that gas-phase chemistry is expected to be dominant. We can safely assume that all hydrocarbons have evaporated from grain surfaces and that there are no ices, since the dust temperature  \citep[100-150\K, ][]{pilleri12a}  is above the sublimation limit for all the molecules  involved in reactions with hydrocarbons, especially methane \citep{tielens05}.

\begin{figure}
\centering
\includegraphics[width = \linewidth]{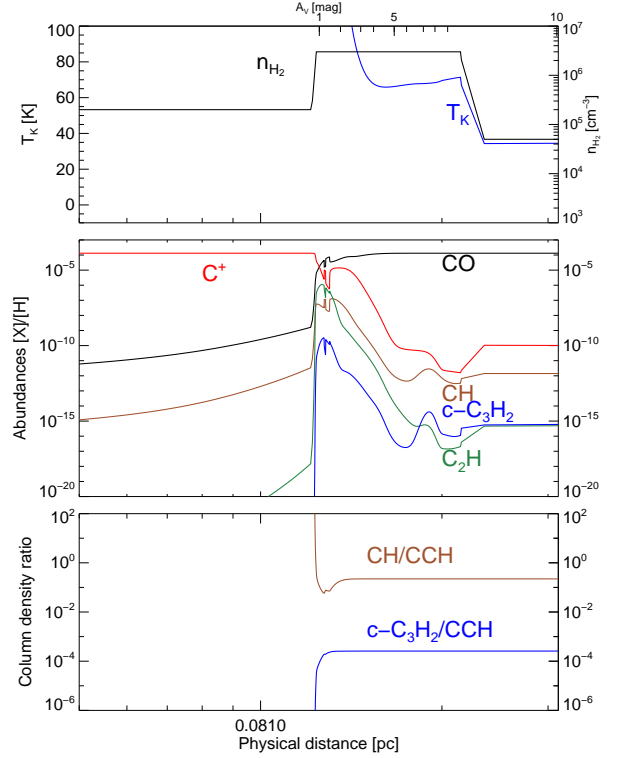}
\caption{ Model results using the gas-phase steady-state chemistry with the Meudon PDR code. The cloud is illuminated from the left side with a UV field of $\hab=\ex{5}{5}$.}
\label{fig_chem}
\end{figure}

  The results of the Meudon  code modeling are displayed in Fig.\,\ref{fig_chem}. The model shows that all the hydrocarbons are expected to have their  abundance peak at  $\Av\sim1$. For \CCH, the model predicts a peak abundance of $\sim10^{-6}$ which rapidly decreases to $10^{-15}$. Similarly, \CthHtw has an abundance peak toward the PDR of $\sim 10^{-10}$ and a decrease to   $10^{-15}$. The abundance  gradient is less abrupt for \CH which passes from a peak of   $10^{-8}$ in the PDR to  $10^{-12}$ in the more shielded layers of the molecular cloud. 

The column densities (and abundances) of \CH and \CCH are relatively well reproduced by the Meudon PDR code, considering the integrated column density in the two layers \Lpdr and \Lhd (see Table \ref{table_comparison}). 
Concerning \CthHtw, the model predicts a \CCr ratio of 0.0003 while the observed value is of the order of 0.01  in the PDR, which shows that the \CthHtw is overabundant compared to the model predictions. This is consistent with several previous studies showing that in PDRs the gas-phase  model predictions  cannot account for  the observed abundances of small hydrocarbons \citep{fuente93, teyssier04, pety05b}, especially \CthHtw.

\subsubsection*{A link with PAHs?}

\label{subsec_linkpah}

A possible way to reduce the discrepancy  could be the injection of  fresh hydrocarbons by the photo-destruction of PAHs  or larger grains \citep{fuente03,pety05}. Indeed, laboratory experiments show that the photodissociation of small PAHs ($N_{\rm C}\le 24$) can lead to the production of small hydrocarbons C$_n$H$_m$ ($n>m$) that can be injected into the gas phase \citep{useli07}. To understand if PAHs can be linked to a chemical differentiation of \CCH and \CthHtw, 
 it is useful to compare the ratio between the column density of  \CthHtw and \CCH with the spatial distribution of  PAHs. 
In Fig.\,\ref{fig_overplot_irac} we display the comparison of the \CCr ratio with the IRAC 8\mum map.

 In low- to mild-UV irradiated PDR, the 8\mum map is dominated by the emission of the AIB at 7.7\mum, but in high-UV excited environments larger grains at thermal equilibrium can reach temperatures high enough to have continuum emission at these wavelength. To check whether this is the case in Mon~R2, we have compared the IRAC 8\mum map with  that obtained with the IRAC 3.5\mum filter, which is expected to be dominated either by the PAH 3.3\mum feature or  by the stellar continuum. The two IRAC maps are very similar, except for very  point-like features in correspondence with the infrared source IRS 2 and IRS 3. Therefore, we can assume that most of the IRAC 8\mum emission is indeed due to AIB emission. Finally, visual inspection of the IRS spectra in the  central region of Mon~R2 shows that the PAH features dominate the emission at 8\mum, except towards the bright infrared sources. 
The comparison of the ratio of the \CthHtw and \CCH column densities with the spatial distribution of \CCH  suggests that $i)$ \CthHtw is overabundant  in the innermost layers of the cloud, and $ii)$ the \CthHtw spatial distribution extends to the NW, following the northern IR ridge that is observed at 8\mum \citep{ginard12}.

The correlation between the PAH emission and the \CthHtw overabundance, although suggestive of a chemical link between these species, does not prove yet that these two tracers are chemically linked (i.e. by UV processing), since they  are both good PDR tracers. For instance, it has been observed that PAH and [\Cp] emission are spatially correlated in PDRs \citep{joblin10}, since both tracers arise in the most exposed layers of the PDR. 
This could also mean that the chemical differentiation of \CCH and \CthHtw in PDRs can also be due to an increased abundance of C$^+$ in these environments. Furthermore, \Cp and PAHs also trace the electron density, which is mainly due to  the ionization of C and photo-electric effect on PAHs.  Since \CthHtw is supposed to form by electron recombination of  the C$_3$H$_3^+$, a higher production efficiency of photo-electrons ejected from PAHs may be a cause of \CthHtw enhancement. However, the inclusion of these effects in PDR models is still very challenging and beyond the scope of this paper.

\subsubsection{Low density envelope}
 Our observations  indicate that the abundances of small hydrocarbons in the envelope are about 5-10 times lower compared to the PDR. The column densities of \CH and \CthHtw relative to \CCH are also a factor of 3 and 2 lower compared to the PDR. Even though, the \CCH and \CthHtw abundances are larger  by several orders of magnitude than those predicted by steady-stage gas chemical models.
In our sketch of MonR2, we assume the envelope to be 40 mag thick  on each side of the spherical cloud, with a gas density of $\nhh = \ex{5}{4}$\cmc and a gas kinetic temperature of 35\K.
With these physical conditions, gas-grain chemistry and time-dependent effects may play a role.

 To study these processes we used the UCL\_CHEM models running
 Phase 1 from diffuse core up to a density of $\nhh= 5\times10^4$\cmc, while the temperature and radiation field were kept at 10\K and 1 Habing respectively. Once the final density was reached we increased the radiation field to $\hab = 100 $ and the kinetic temperature to 35\K (Phase 2).  The model assumes a number of parameters, such as the freeze-out,  photo- and thermal-desorption efficiencies.  We use the standard values for the different non thermal desorption efficiencies 
as in \citet{roberts07}. In particular, the models are strongly dependent on the freeze-out efficiency during phase 1, which determines the initial abundances of \CH, \CCH and \CthHtw\  in Phase 2.  
  We have tested the impact of the kinetic temperature in the envelope in the final hydrocarbon abundances, and found no significant difference in the range [10-35]\K. This is consistent with the fact that at $T_{\rm dust}$=35\K, most of the species remain in the grain mantles 
\citep{viti04} and the rates of main gas phase reactions involved in the formation and 
destruction of hydrocarbons do not present strong variations in this temperature range.

\begin{figure}[ht]
\centering
\sidecaption
\includegraphics[width=1\linewidth]{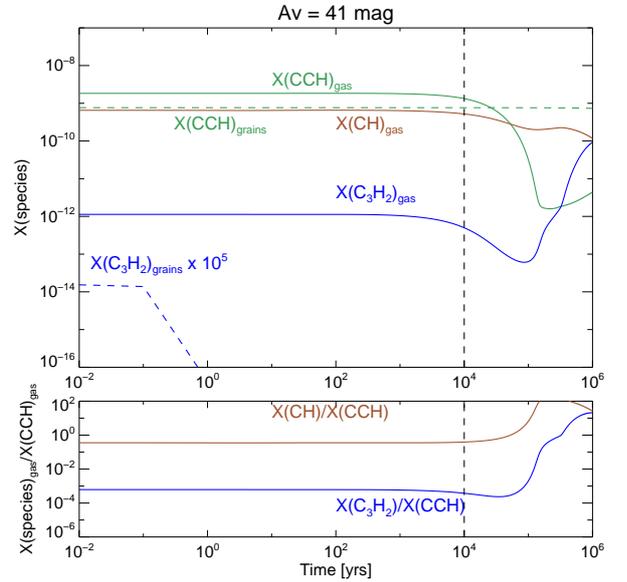}
\caption{Results of the UCL CHEM code to model the low-density envelope \Lhd\  for a starting CO freeze-out  of 10\%. The results are given at   \Av of 41 magnitudes and by setting the temperature fixed to 35\K.   The bottom panel shows the  ratio of gas-phase \CH and \CthHtw relative to \CCH. The \bf vertical dashed  line marks the  age of $10^4$\,years,  which is used in Table \ref{table_comparison}}. 
\label{fig_uclchem3}
\end{figure}

In Fig. \ref{fig_uclchem3} we report an example of the time evolution during Phase 2 for an $\Av = 41$, assuming a low (10\%) CO freeze-out. 
 The model predicts that all the hydrocarbon abundances are constant up to $10^4$ years. At this time, each of the absolute abundances for \CH and \CCH are reproduced within a factor of 3
 but their ratio is over-predicted by a factor of  7 (see Table \ref{table_comparison}). Considering the high uncertainties in the absolute column density of \CH (driven by the unknown excitation temperature) it is difficult to further constrain the model based on this ratio. After $\ex{1}{4}$ years,  gas phase \CH decreases by a factor of a few, whereas \CCH and \CthHtw decrease by the more than one order of magnitude: for $t\gtrsim10^5$ years, the agreement of \CCr\  with observations improves, but the absolute abundances and the \CH/\CCH ratio do not agree as well as at $\ex{1}{4}$ years.
  
 Repeating the calculations for higher freeze-out  fractions (25\% and 45\%, see Table \ref{table_comparison}) one can find a better agreement for one or the other molecule, but it is difficult to find a solution that fits reasonably well all the abundances or column densities.  

In any case, the chemistry of Mon~R2 seems to be dominated by grain-surface   processes 
and time-dependent effects, since at the age of UCHII region ($\sim$0.1 Myr) the 
molecular chemistry is far from the steady state yet. In this scenario  the 
current abundances of small hydrocarbons are dependent on the previous history 
of the cloud. This is consistent with recent results of \citet{mookerjea12}, who studied these effects in the envelope associated with DR21. They investigated the impact of time-dependent effects for  the carbon chemistry of this warm cloud, showing that PAH-related chemistry is not needed to explain the abundances of neither \CCH or \CthHtw in this molecular envelope. 

\begin{table*}
\begin{center}
\caption{Comparison between the abundances derived from observations and model predictions. For the UCL\_CHEM code, we show the results at $t=10^4$ years. }
\label{table_comparison}
\begin{tabular}{lcccccc}
\toprule
			&	\Av	&			$X$(\CH)	\tablefootmark{(a)}				&	$X$(\CCH	)		\tablefootmark{(a)}		&	$X$(\CthHtw)		\tablefootmark{(a)}		&	$N(\CH)/N(\CCH)$ 		&	\CCr		\\
			&	[mag]&&&\\
\midrule
$X_{obs}^{PDR}$&1-10&$\ex{3.3^{+6.0}_{-2.2}}{-9}$	& 	\ex{1.7}{-8}		&	\ex{1.4}{-10}	 	& 0.15			& 0.008 \vspace{0.1cm}\\
Meudon \Lpdr+\Lhd		&1-10		& \ex{1.1}{-8} &	 \ex{4.8}{-8}  	&	 \ex{1.3}{-11}  & 0.22 & 0.00025	\\

\midrule
$X_{obs}^{env}$	&	10-50		&\ex{1.9^{+3.3}_{-1.2}}{-10}	& 	\ex{3.5}{-9}		&	\ex{1.4}{-11}		& 0.054			& 0.004	\\	

UCL\_CHEM \Lenv	f = 10\tablefootmark{(b)} & 41 		& \ex{5.1}{-10}				&	\ex{1.3}{-9}		 &	\ex{4.8}{-13}	& 0.39		& 0.0004\\	
UCL\_CHEM \Lenv	f = 25\tablefootmark{(b)} & 41 		& \ex{9.5}{-10}				&	\ex{6.9}{-10}		 &	\ex{3.5}{-11}	& 1.37		& 0.05\\	
UCL\_CHEM \Lenv	f = 45\tablefootmark{(b)} & 41 		& \ex{7.5}{-10}				&	\ex{3.5}{-10}		 &	\ex{1.4}{-10}	& 2.1		& 0.40\\	
\bottomrule
\end{tabular}
\tablefoot{For the observed abundances and the Meudon PDR results, we report mean values of the abundances, while for the UCL\_CHEM results, we report the value at a given \Av, and assume that these values are relatively constant in the respective layers.\\ $^{(a)}$ Except for \CH, where the uncertainty is driven by the excitation temperature, the abundance uncertainty is driven by the \HH gas reference, and is a factor of 2. \\ $^{(b)}$ The $f$ value indicates the percentage of CO on the icy mantles at the end of Phase I.}
\end{center}
\end{table*}

\section{Conclusions}
\label{sec_conclusions}

In this paper, we have presented new observations of small hydrocarbon molecules (\CH, \CCH and \CthHtw) in MonR2 obtained with the \IRAMthm telescope and the HIFI instrument onboard \herschel.  In particular, the HIFI observations allowed to access the  N=6-5 rotational transitions of \CCH with a good spatial and spectral resolution. This allowed to constrain the excitation conditions and to distinguish the different phases in which this species is abundant.
Profiting of a multi-transition and spatial study, we have derived the spatial distribution of these molecules, determining  their abundances in the different layers of Mon~R2. 

We compared the observational results with  predictions  of different PDR models, to investigate which kind of chemistry can dominate in the different environments that co-exist in MonR2: 
\begin{itemize}
\item In the high-UV illuminated PDR, grain mantles have already been evaporated and we can safely assume that chemical equilibrium has  been reached.  In this case, the gas-phase chemistry can explain quite well the abundances of \CH and \CCH but fails to reproduce \CthHtw. The chemical differentiation of these two molecules  happens in correspondence with PAH emission, suggesting once again a link between these species. 

\item In the cold ($\sim 35\K$) low-density molecular cloud, time dependent effects seems to be very important: around $10^4$ years, the abundances predicted by the models are relatively consistent with those derived from the observations.  At this time all the absolute abundances are  relatively well reproduced, although \CH is over-predicted relative to both \CCH and \CthHtw. These models are however strongly dependent on the initial abundances due to the freeze-out in the cloud collapse phase.
\end{itemize}

This study shows that several effects must be taken into account to understand the chemistry of small hydrocarbons such as \CCH and \CthHtw: the relative contribution of UV radiation, grain-surface 
chemistry 
and time-dependent effects may vary in different environments. Regarding 
PDRs, a more detailed analysis of the small hydrocarbon chemistry in these 
 environments requires: $(i)$ a larger sample of PDRs spanning a large range 
of physical conditions  to minimize effects due to 
a peculiar geometry of a particular source;  $(ii)$  to revise the possible chemical routes to the formation of small hydrocarbons such as the photo-destruction of PAHs,
 and very small grains. 
\herschel data will allow to constrain the hydrocarbons chemical models, including related species that are not observable from the ground (e.g., C$^+$, \CHp). Finally ($iii$) higher resolution observations with the Plateau de Bure Interferometer, NOEMA, ALMA and JCMT  will allow to observe  several key species with an unprecedented spatial resolution. This will enable to resolve the small-scale chemical frontiers in the PDR that seems to be key interpret the link between small hydrocarbons and PAHs.

\begin{acknowledgements}
 The authors thank the referee for his useful comments. \\
HIFI has been designed and built by a consortium of institutes and university departments from across
Europe, Canada and the United States under the leadership of SRON Netherlands Institute for Space
Research, Groningen, The Netherlands and with major contributions from Germany, France and the US.
Consortium members are: Canada: CSA, U.Waterloo; France: CESR, LAB, LERMA, IRAM; Germany:
KOSMA, MPIfR, MPS; Ireland, NUI Maynooth; Italy: ASI, IFSI-INAF, Osservatorio Astrofisico di Arcetri-
INAF; Netherlands: SRON, TUD; Poland: CAMK, CBK; Spain: Observatorio Astron\'omico Nacional (IGN),
Centro de Astrobiolog\'{\i}a (CSIC-INTA). Sweden: Chalmers University of Technology - MC2, RSS \& GARD;
Onsala Space Observatory; Swedish National Space Board, Stockholm University - Stockholm Observatory;
Switzerland: ETH Zurich, FHNW; USA: Caltech, JPL, NHSC. \\
This paper was partially supported by Spanish MICINN
under project AYA2009-07304 and within the program CONSOLIDER INGENIO 2010, under grant Molecular Astrophysics: 
The Herschel and ALMA Era ASTROMOL (ref.: CSD2009-00038). \\
French scientists are supported by CNES for the
Herschel results.\\
Part of this work was supported by the Deutsche
Forschungsgemeinschaft, project number SFB 956 C1.\\
JRG is supported by a Ramo\'n y Cajal research contract from the MINECO and co-financed by the European Social Fund.

\end{acknowledgements}

\bibliographystyle{aa}
\bibliography{biblio}

\begin{thebibliography}{48}
\expandafter\ifx\csname natexlab\endcsname\relax\def\natexlab#1{#1}\fi

\bibitem[{{Bern{\'e}} {et~al.}(2009){Bern{\'e}}, {Fuente}, {Goicoechea},
  {Pilleri}, {Gonz{\'a}lez-Garc{\'{\i}}a}, \& {Joblin}}]{berne09a}
{Bern{\'e}}, O., {Fuente}, A., {Goicoechea}, J.~R., {et~al.} 2009, \apjl, 706,
  L160

\bibitem[{{Cernicharo}(2012)}]{cernicharo12}
{Cernicharo}, J. 2012, in Proceedings of the European Conference on Laboratory
  Astrophysics, Eur. Astron. Soc. Publ. Ser, ed. C.~{Stehl\'e}, C.~{Joblin}, \&
  L.~{d'Hendecourt}

\bibitem[{{Chandra} \& {Kegel}(2000)}]{chandra00}
{Chandra}, S. \& {Kegel}, W.~H. 2000, \aaps, 142, 113

\bibitem[{{Choi} {et~al.}(2000){Choi}, {Evans}, {Tafalla}, \&
  {Bachiller}}]{choi00}
{Choi}, M., {Evans}, II, N.~J., {Tafalla}, M., \& {Bachiller}, R. 2000, ApJ,
  538, 738

\bibitem[{{de Graauw} {et~al.}(2010){de Graauw}, {Helmich}, {Phillips},
  {Stutzki}, {Caux}, {Whyborn}, {Dieleman}, {Roelfsema}, {Aarts}, {Assendorp},
  {Bachiller}, {Baechtold}, {Barcia}, {Beintema}, {Belitsky}, {Benz}, {Bieber},
  {Boogert}, {Borys}, {Bumble}, {Ca{\"i}s}, {Caris}, {Cerulli-Irelli},
  {Chattopadhyay}, {Cherednichenko}, {Ciechanowicz}, {Coeur-Joly}, {Comito},
  {Cros}, {de Jonge}, {de Lange}, {Delforges}, {Delorme}, {den Boggende},
  {Desbat}, {Diez-Gonz{\'a}lez}, {di Giorgio}, {Dubbeldam}, {Edwards},
  {Eggens}, {Erickson}, {Evers}, {Fich}, {Finn}, {Franke}, {Gaier}, {Gal},
  {Gao}, {Gallego}, {Gauffre}, {Gill}, {Glenz}, {Golstein}, {Goulooze},
  {Gunsing}, {G{\"u}sten}, {Hartogh}, {Hatch}, {Higgins}, {Honingh}, {Huisman},
  {Jackson}, {Jacobs}, {Jacobs}, {Jarchow}, {Javadi}, {Jellema}, {Justen},
  {Karpov}, {Kasemann}, {Kawamura}, {Keizer}, {Kester}, {Klapwijk}, {Klein},
  {Kollberg}, {Kooi}, {Kooiman}, {Kopf}, {Krause}, {Krieg}, {Kramer},
  {Kruizenga}, {Kuhn}, {Laauwen}, {Lai}, {Larsson}, {Leduc}, {Leinz}, {Lin},
  {Liseau}, {Liu}, {Loose}, {L{\'o}pez-Fernandez}, {Lord}, {Luinge}, {Marston},
  {Mart{\'{\i}}n-Pintado}, {Maestrini}, {Maiwald}, {McCoey}, {Mehdi}, {Megej},
  {Melchior}, {Meinsma}, {Merkel}, {Michalska}, {Monstein}, {Moratschke},
  {Morris}, {Muller}, {Murphy}, {Naber}, {Natale}, {Nowosielski}, {Nuzzolo},
  {Olberg}, {Olbrich}, {Orfei}, {Orleanski}, {Ossenkopf}, {Peacock}, {Pearson},
  {Peron}, {Phillip-May}, {Piazzo}, {Planesas}, {Rataj}, {Ravera}, {Risacher},
  {Salez}, {Samoska}, {Saraceno}, {Schieder}, {Schlecht}, {Schl{\"o}der},
  {Schm{\"u}lling}, {Schultz}, {Schuster}, {Siebertz}, {Smit}, {Szczerba},
  {Shipman}, {Steinmetz}, {Stern}, {Stokroos}, {Teipen}, {Teyssier}, {Tils},
  {Trappe}, {van Baaren}, {van Leeuwen}, {van de Stadt}, {Visser}, {Wildeman},
  {Wafelbakker}, {Ward}, {Wesselius}, {Wild}, {Wulff}, {Wunsch}, {Tielens},
  {Zaal}, {Zirath}, {Zmuidzinas}, \& {Zwart}}]{degraauw10}
{de Graauw}, T., {Helmich}, F.~P., {Phillips}, T.~G., {et~al.} 2010, \aap, 518,
  L6

\bibitem[{{Fontani} {et~al.}(2012){Fontani}, {Palau}, {Busquet}, {Isella},
  {Estalella}, {Sanchez-Monge}, {Caselli}, \& {Zhang}}]{fontani12}
{Fontani}, F., {Palau}, A., {Busquet}, G., {et~al.} 2012, \mnras, 423, 1691

\bibitem[{{Fuente} {et~al.}(2010){Fuente}, {Bern{\'e}}, {Cernicharo}, {Rizzo},
  {Gonz{\'a}lez-Garc{\'{\i}}a}, {Goicoechea}, {Pilleri}, {Ossenkopf}, {Gerin},
  {G{\"u}sten}, {Akyilmaz}, {Benz}, {Boulanger}, {Bruderer}, {Dedes}, {France},
  {Garc{\'{\i}}a-Burillo}, {Harris}, {Joblin}, {Klein}, {Kramer}, {Le Petit},
  {Lord}, {Martin}, {Mart{\'{\i}}n-Pintado}, {Mookerjea}, {Neufeld}, {Okada},
  {Pety}, {Phillips}, {R{\"o}llig}, {Simon}, {Stutzki}, {van der Tak},
  {Teyssier}, {Usero}, {Yorke}, {Schuster}, {Melchior}, {Lorenzani},
  {Szczerba}, {Fich}, {McCoey}, {Pearson}, \& {Dieleman}}]{fuente10}
{Fuente}, A., {Bern{\'e}}, O., {Cernicharo}, J., {et~al.} 2010, \aap, 521, L23

\bibitem[{{Fuente} {et~al.}(1993){Fuente}, {Martin-Pintado}, {Cernicharo}, \&
  {Bachiller}}]{fuente93}
{Fuente}, A., {Martin-Pintado}, J., {Cernicharo}, J., \& {Bachiller}, R. 1993,
  \aap, 276, 473

\bibitem[{{Fuente} {et~al.}(2003){Fuente}, {Rodr{\i}guez-Franco},
  {Garc{\i}a-Burillo}, {Mart{\i}n-Pintado}, \& {Black}}]{fuente03}
{Fuente}, A., {Rodr{\i}guez-Franco}, A., {Garc{\i}a-Burillo}, S.,
  {Mart{\i}n-Pintado}, J., \& {Black}, J.~H. 2003, \aap, 406, 899

\bibitem[{{Fuente} {et~al.}(1996){Fuente}, {Rodriguez-Franco}, \&
  {Martin-Pintado}}]{fuente96}
{Fuente}, A., {Rodriguez-Franco}, A., \& {Martin-Pintado}, J. 1996, \aap, 312,
  599

\bibitem[{{Gerin} {et~al.}(2011){Gerin}, {Ka{\'z}mierczak}, {Jastrzebska},
  {Falgarone}, {Hily-Blant}, {Godard}, \& {de Luca}}]{gerin11}
{Gerin}, M., {Ka{\'z}mierczak}, M., {Jastrzebska}, M., {et~al.} 2011, \aap,
  525, A116

\bibitem[{{Ginard} {et~al.}(2012){Ginard}, {Gonz{\'a}lez-Garc{\'{\i}}a},
  {Fuente}, {Cernicharo}, {Alonso-Albi}, {Pilleri}, {Gerin},
  {Garc{\'{\i}}a-Burillo}, {Ossenkopf}, {Rizzo}, {Kramer}, {Goicoechea},
  {Pety}, {Bern{\'e}}, \& {Joblin}}]{ginard12}
{Ginard}, D., {Gonz{\'a}lez-Garc{\'{\i}}a}, M., {Fuente}, A., {et~al.} 2012,
  \aap, 543, A27

\bibitem[{{Goicoechea} \& {Le Bourlot}(2007)}]{goicoechea07}
{Goicoechea}, J.~R. \& {Le Bourlot}, J. 2007, \aap, 467, 1

\bibitem[{{Gonzalez Garcia} {et~al.}(2008){Gonzalez Garcia}, {Le Bourlot}, {Le
  Petit}, \& {Roueff}}]{gonzalez08}
{Gonzalez Garcia}, M., {Le Bourlot}, J., {Le Petit}, F., \& {Roueff}, E. 2008,
  \aap, 485, 127

\bibitem[{{Habing}(1968)}]{habing68}
{Habing}, H.~J. 1968, \bain, 19, 421

\bibitem[{{Henning} {et~al.}(1992){Henning}, {Chini}, \& {Pfau}}]{henning92}
{Henning}, T., {Chini}, R., \& {Pfau}, W. 1992, \aap, 263, 285

\bibitem[{{Joblin} {et~al.}(2010){Joblin}, {Pilleri}, {Montillaud}, {Fuente},
  {Gerin}, {Bern{\'e}}, {Ossenkopf}, {Le Bourlot}, {Teyssier}, {Goicoechea},
  {Le Petit}, {R{\"o}llig}, {Akyilmaz}, {Benz}, {Boulanger}, {Bruderer},
  {Dedes}, {France}, {G{\"u}sten}, {Harris}, {Klein}, {Kramer}, {Lord},
  {Martin}, {Martin-Pintado}, {Mookerjea}, {Okada}, {Phillips}, {Rizzo},
  {Simon}, {Stutzki}, {van der Tak}, {Yorke}, {Steinmetz}, {Jarchow},
  {Hartogh}, {Honingh}, {Siebertz}, {Caux}, \& {Colin}}]{joblin10}
{Joblin}, C., {Pilleri}, P., {Montillaud}, J., {et~al.} 2010, \aap, 521, L25

\bibitem[{{Le Bourlot} {et~al.}(2012){Le Bourlot}, {Le Petit}, {Pinto},
  {Roueff}, \& {Roy}}]{lebourlot12}
{Le Bourlot}, J., {Le Petit}, F., {Pinto}, C., {Roueff}, E., \& {Roy}, F. 2012,
  \aap, 541, A76

\bibitem[{{Le Petit} {et~al.}(2006){Le Petit}, {Nehm{\'e}}, {Le Bourlot}, \&
  {Roueff}}]{lepetit06}
{Le Petit}, F., {Nehm{\'e}}, C., {Le Bourlot}, J., \& {Roueff}, E. 2006, \apjs,
  164, 506

\bibitem[{{Liszt} \& {Lucas}(2002)}]{liszt02}
{Liszt}, H. \& {Lucas}, R. 2002, \aap, 391, 693

\bibitem[{{Loren}(1977)}]{loren77}
{Loren}, R.~B. 1977, \apj, 215, 129

\bibitem[{{Lucas} \& {Liszt}(2000)}]{lucas00}
{Lucas}, R. \& {Liszt}, H.~S. 2000, \aap, 358, 1069

\bibitem[{{McCarthy} {et~al.}(2006){McCarthy}, {Mohamed}, {Brown}, \&
  {Thaddeus}}]{mccarthy06}
{McCarthy}, M.~C., {Mohamed}, S., {Brown}, J.~M., \& {Thaddeus}, P. 2006,
  Proceedings of the National Academy of Science, 103, 12263

\bibitem[{{Mookerjea} {et~al.}(2012){Mookerjea}, {Hassel}, {Gerin}, {Giesen},
  {Stutzki}, {Herbst}, {Black}, {Goldsmith}, {Menten}, {Kre{\l}owski}, {De
  Luca}, {Csengeri}, {Joblin}, {Ka{\'z}mierczak}, {Schmidt}, {Goicoechea}, \&
  {Cernicharo}}]{mookerjea12}
{Mookerjea}, B., {Hassel}, G.~E., {Gerin}, M., {et~al.} 2012, \aap, 546, A75

\bibitem[{{Ossenkopf} {et~al.}(2011){Ossenkopf}, {R{\"o}llig}, {Kramer},
  {Okada}, {Fuente}, {Akyilmaz Yabaci}, {Benz}, {Bern{\'e}}, {Boulanger},
  {Bruderer}, {Dedes}, {France}, {Gerin}, {Goicoechea}, {Gusdorf},
  {G{\"u}sten}, {Harris}, {Joblin}, {Klein}, {Latter}, {Le Petit}, {Lord},
  {Martin}, {Pilleri}, {Martin-Pintado}, {Mookerjea}, {Neufeld}, {Phillips},
  {Rizzo}, {Simon}, {Stutzki}, {van der Tak}, {Teyssier}, \&
  {Yorke}}]{ossenkopf11}
{Ossenkopf}, V., {R{\"o}llig}, M., {Kramer}, C., {et~al.} 2011, in EAS
  Publications Series, Vol.~52, EAS Publications Series, ed. M.~{R{\"o}llig},
  R.~{Simon}, V.~{Ossenkopf}, \& J.~{Stutzki}, 181--186

\bibitem[{{Ossenkopf} {et~al.}(2013){Ossenkopf}, {R{\"o}llig}, {Neufeld},
  {Pilleri}, {Lis}, {Fuente}, {van der Tak}, \& {Bergin}}]{ossenkopf13}
{Ossenkopf}, V., {R{\"o}llig}, M., {Neufeld}, D.~A., {et~al.} 2013, \aap, 550,
  A57

\bibitem[{{Ott}(2010)}]{ott10}
{Ott}, S. 2010, in Astronomical Society of the Pacific Conference Series, Vol.
  434, Astronomical Data Analysis Software and Systems XIX, ed. {Y.~Mizumoto,
  K.-I.~Morita, \& M.~Ohishi}, 139

\bibitem[{{Padovani} {et~al.}(2009){Padovani}, {Walmsley}, {Tafalla}, {Galli},
  \& {M{\"u}ller}}]{padovani09}
{Padovani}, M., {Walmsley}, C.~M., {Tafalla}, M., {Galli}, D., \& {M{\"u}ller},
  H.~S.~P. 2009, \aap, 505, 1199

\bibitem[{{Penzias} \& {Burrus}(1973)}]{penzias73}
{Penzias}, A.~A. \& {Burrus}, C.~A. 1973, \araa, 11, 51

\bibitem[{{Pety}(2005)}]{pety05}
{Pety}, J. 2005, in SF2A-2005: Semaine de l'Astrophysique Francaise, ed.
  {F.~Casoli, T.~Contini, J.~M.~Hameury, \& L.~Pagani}, 721

\bibitem[{{Pety} {et~al.}(2005){Pety}, {Teyssier}, {Foss{\'e}}, {Gerin},
  {Roueff}, {Abergel}, {Habart}, \& {Cernicharo}}]{pety05b}
{Pety}, J., {Teyssier}, D., {Foss{\'e}}, D., {et~al.} 2005, \aap, 435, 885

\bibitem[{{Pilbratt} {et~al.}(2010){Pilbratt}, {Riedinger}, {Passvogel},
  {Crone}, {Doyle}, {Gageur}, {Heras}, {Jewell}, {Metcalfe}, {Ott}, \&
  {Schmidt}}]{pilbratt10}
{Pilbratt}, G.~L., {Riedinger}, J.~R., {Passvogel}, T., {et~al.} 2010, \aap,
  518, L1

\bibitem[{{Pilleri} {et~al.}(2012{\natexlab{a}}){Pilleri}, {Fuente},
  {Cernicharo}, {Ossenkopf}, {Bern{\'e}}, {Gerin}, {Pety}, {Goicoechea},
  {Rizzo}, {Montillaud}, {Gonz{\'a}lez-Garc{\'{\i}}a}, {Joblin}, {Le Bourlot},
  {Le Petit}, \& {Kramer}}]{pilleri12a}
{Pilleri}, P., {Fuente}, A., {Cernicharo}, J., {et~al.} 2012{\natexlab{a}},
  \aap, 544, A110

\bibitem[{{Pilleri} {et~al.}(2012{\natexlab{b}}){Pilleri}, {Montillaud},
  {Bern{\'e}}, \& {Joblin}}]{pilleri12b}
{Pilleri}, P., {Montillaud}, J., {Bern{\'e}}, O., \& {Joblin}, C.
  2012{\natexlab{b}}, \aap, 542, A69

\bibitem[{{Rizzo} {et~al.}(2003){Rizzo}, {Fuente}, {Rodr{\'{\i}}guez-Franco},
  \& {Garc{\'{\i}}a-Burillo}}]{rizzo03}
{Rizzo}, J.~R., {Fuente}, A., {Rodr{\'{\i}}guez-Franco}, A., \&
  {Garc{\'{\i}}a-Burillo}, S. 2003, ApJ, 597, L153

\bibitem[{{Roberts} {et~al.}(2007){Roberts}, {Rawlings}, {Viti}, \&
  {Williams}}]{roberts07}
{Roberts}, J.~F., {Rawlings}, J.~M.~C., {Viti}, S., \& {Williams}, D.~A. 2007,
  \mnras, 382, 733

\bibitem[{{Roelfsema} {et~al.}(2012){Roelfsema}, {Helmich}, {Teyssier},
  {Ossenkopf}, {Morris}, {Olberg}, {Shipman}, {Risacher}, {Akyilmaz},
  {Assendorp}, {Avruch}, {Beintema}, {Biver}, {Boogert}, {Borys}, {Braine},
  {Caris}, {Caux}, {Cernicharo}, {Coeur-Joly}, {Comito}, {de Lange},
  {Delforge}, {Dieleman}, {Dubbeldam}, {de Graauw}, {Edwards}, {Fich},
  {Flederus}, {Gal}, {di Giorgio}, {Herpin}, {Higgins}, {Hoac}, {Huisman},
  {Jarchow}, {Jellema}, {de Jonge}, {Kester}, {Klein}, {Kooi}, {Kramer},
  {Laauwen}, {Larsson}, {Leinz}, {Lord}, {Lorenzani}, {Luinge}, {Marston},
  {Mart{\'{\i}}n-Pintado}, {McCoey}, {Melchior}, {Michalska}, {Moreno},
  {M{\"u}ller}, {Nowosielski}, {Okada}, {Orlea{\'n}ski}, {Phillips}, {Pearson},
  {Rabois}, {Ravera}, {Rector}, {Rengel}, {Sagawa}, {Salomons},
  {S{\'a}nchez-Su{\'a}rez}, {Schieder}, {Schl{\"o}der}, {Schm{\"u}lling},
  {Soldati}, {Stutzki}, {Thomas}, {Tielens}, {Vastel}, {Wildeman}, {Xie},
  {Xilouris}, {Wafelbakker}, {Whyborn}, {Zaal}, {Bell}, {Bjerkeli}, {De Beck},
  {Cavali{\'e}}, {Crockett}, {Hily-Blant}, {Kama}, {Kaminski}, {Lefl{\'o}ch},
  {Lombaert}, {de Luca}, {Makai}, {Marseille}, {Nagy}, {Pacheco}, {van der
  Wiel}, {Wang}, \& {Y{\i}ld{\i}z}}]{roelfsema12}
{Roelfsema}, P.~R., {Helmich}, F.~P., {Teyssier}, D., {et~al.} 2012, \aap, 537,
  A17

\bibitem[{{Sheffer} {et~al.}(2008){Sheffer}, {Rogers}, {Federman}, {Abel},
  {Gredel}, {Lambert}, \& {Shaw}}]{sheffer08}
{Sheffer}, Y., {Rogers}, M., {Federman}, S.~R., {et~al.} 2008, \apj, 687, 1075

\bibitem[{{Spielfiedel} {et~al.}(2012){Spielfiedel}, {Feautrier}, {Najar}, {Ben
  Abdallah}, {Dayou}, {Senent}, \& {Lique}}]{spielfiedel12}
{Spielfiedel}, A., {Feautrier}, N., {Najar}, F., {et~al.} 2012, \mnras, 421,
  1891

\bibitem[{{Tafalla} {et~al.}(1994){Tafalla}, {Bachiller}, \&
  {Wright}}]{tafalla94}
{Tafalla}, M., {Bachiller}, R., \& {Wright}, M.~C.~H. 1994, \apjl, 432, L127

\bibitem[{{Tafalla} {et~al.}(1997){Tafalla}, {Bachiller}, {Wright}, \&
  {Welch}}]{tafalla97}
{Tafalla}, M., {Bachiller}, R., {Wright}, M.~C.~H., \& {Welch}, W.~J. 1997,
  \apj, 474, 329

\bibitem[{{Teyssier} {et~al.}(2004){Teyssier}, {Foss{\'e}}, {Gerin}, {Pety},
  {Abergel}, \& {Roueff}}]{teyssier04}
{Teyssier}, D., {Foss{\'e}}, D., {Gerin}, M., {et~al.} 2004, \aap, 417, 135

\bibitem[{{Tielens}(2005)}]{tielens05}
{Tielens}, A.~G.~G.~M. 2005, {The Physics and Chemistry of the Interstellar
  Medium}

\bibitem[{{Trevi\~no-Morales, et al.}(2012)}]{trevino12}
{Trevi\~no-Morales, et al.} 2012, \aap, in preparation

\bibitem[{{Ungerechts} {et~al.}(1997){Ungerechts}, {Bergin}, {Goldsmith},
  {Irvine}, {Schloerb}, \& {Snell}}]{ungerechts97}
{Ungerechts}, H., {Bergin}, E.~A., {Goldsmith}, P.~F., {et~al.} 1997, \apj,
  482, 245

\bibitem[{{Useli Bacchitta} \& {Joblin}(2007)}]{useli07}
{Useli Bacchitta}, F. \& {Joblin}, C. 2007, in Molecules in Space and
  Laboratory, ed. P.~J.L. Lemaire \& F.~Combes, publ. S.~Diana, 89

\bibitem[{{Viti} {et~al.}(2004){Viti}, {Collings}, {Dever}, {McCoustra}, \&
  {Williams}}]{viti04}
{Viti}, S., {Collings}, M.~P., {Dever}, J.~W., {McCoustra}, M.~R.~S., \&
  {Williams}, D.~A. 2004, \mnras, 354, 1141

\bibitem[{{Viti} {et~al.}(2011){Viti}, {Jimenez-Serra}, {Yates}, {Codella},
  {Vasta}, {Caselli}, {Lefloch}, \& {Ceccarelli}}]{viti11}
{Viti}, S., {Jimenez-Serra}, I., {Yates}, J.~A., {et~al.} 2011, \apjl, 740, L3

\end{thebibliography}
 \newpage
 \begin{appendix}
  \section{Diffuse gas in the line of sight}
\label{appen_abs}
The low-lying transitions of all the hydrocarbons  
show a clear self-absorption at about the velocity of $\sim$11\kms toward the IF (see Fig. \ref{fig_obsiram}). A similar, narrow self-absorption component has also been reported for the [\Cp] 158\mum line \citep{pilleri12a, ossenkopf13}. 
This feature suggests the presence of a relative diffuse phase in the line of sight. The velocity at which the absorption features appears is the same as that of the emission of the quiescent molecular cloud. Thus, it is likely that this absorbing layer is a low-density gas associated with the most external layer of the molecular cloud. 

To determine the column density of the absorbing gas, we need to estimate a reference for the spectral profile of the emission line. We assume that the widths of the lines belonging to a species are similar, and use a transition that is less affected by self-absorption to estimate the FWHM of a gaussian profile. We obtain a FWHM of 5.5\kms for \CCH (using the J = 87.407\GHz line) and 3\kms for \CthHtw (using the 150.820\GHz line) and  6\kms for \CH. 
Using these line widths, we can adjust the blue-shifted wing of the self-absorbed line with a gaussian and use it as reference to calculate the integrated opacity of the self-absorption. 
By assuming an excitation temperature $T_{ex} =  2.73\K$, we derive column densities associated to the foreground of 
$N(\CH) =		 \ex{4.8}{13}$\cmq
$N(\CCH) =    	\ex{5.0}{13}$\cmq
and $N(\CthHtw) =    \ex{1.4}{12}\cmq. $

From CH, we can derive the \HH column density associated with the
absorption, hence an estimate of the extinction.
The standard abundance of \CH relative to \HH in diffuse gas is \ex{3.5}{-8} \citep{sheffer08}, which leads to a molecular hydrogen column density of $N(\HH) \sim \ex{1.3}{21}$\cmq or about 1.3 mag of extinction. This yields \CCH and \CthHtw abundances (relative to \HH)  of \ex{3.9}{-8} and \ex{1.6}{-9}, respectively. These values are consistent with the typical values observed in diffuse gas \citep{liszt02, gerin11}. The results are similar when considering an offset of [20\arcsec,20\arcsec], where the self-absorption is still detected with a sufficient S/N to infer relatively accurate column densities.  

These estimates suffer from the uncertainty due to the assumed line profile and on the observed continuum. Nevertheless, the detection of this foreground, 1\Av thick layer associated with the external layer of the envelope is interesting and moves toward a deeper understanding of this source. 

\newpage
 \section{Spectroscopic parameters, maps and rotational diagrams}
 
\begin{table}[ht]
{\scriptsize
\caption{ Spectroscopic parameters of the observed transitions}
\label{tab:spec}
\begin{tabular}{lccccl}
\toprule
Transition & Frequency          & E$_u$  	& A$_{ul}$ 	    & $g_u$ \\
	  	&  [\GHz]    	& $[$K$]$  & [s$^{-1}$] & \\
\midrule
\multicolumn{2}{l}{\CH $2\Pi_{1/2}$, $N = 1$, $J = 3/2 - 1/2$} \\
\hspace{0.2cm} P = $2^--1^+$      & 536.76115  & 0.00072  & 0.0638	& 5 \\
\hspace{0.2cm} P = $1^--1^+$       & 536.78195  & 0.00072  & 0.0213	& 3 \\
\hspace{0.2cm} P =  $1^--0^+$       & 536.79568 & 0.0     & 0.0425	& 3 \\
\midrule
\multicolumn{2}{l}{\CH $2\Pi_{1/2}$, $N = 2$, $J = 5/2 - 3/2$} \\
\hspace{0.2cm} P = $3^+-2^-$       &1656.961  &	25.727  & 0.03719	& 7 \\
\hspace{0.2cm} P = $2^+-2^-$       &  1656.970 &	25.727  & 0.00372	& 5 \\
\hspace{0.2cm} P =  $2^+-1^-$       & 1656.972  &	25.727   & 0.03348	& 5 \\

\midrule
\multicolumn{2}{l}{\CCH $N = 1-0$}\\ 
\hspace{0.2cm}J=3/2-1/2, F= 1-1 	& 	87.28410  &		0.00216  &	\ex{2.5920}{-7}& 3\\
\hspace{0.2cm}J=3/2-1/2, F= 2-1 	&	87.31690  &		0.00216  &	\ex{1.5274}{-6}& 5\\
\hspace{0.2cm}J=3/2-1/2, F= 1-0 	&	87.32859  &		0.00000  &	\ex{1.2684}{-6}& 3\\
\hspace{0.2cm}J=1/2-1/2, F= 1-1 	&	87.40199  &		0.00216  &	\ex{1.2715}{-6}& 3\\
\hspace{0.2cm}J=1/2-1/2, F= 0-1 	&	87.40716  &		0.00216  &	\ex{1.5324}{-6}& 1\\
\midrule
\multicolumn{2}{l}{\CCH $N = 3-2$}\\ 
\hspace{0.2cm}J=7/2-5/2, F=4-3 	& 	262.00426	 &  	12.57513  &  \ex{5.30542}{-5}	& 9\\
\hspace{0.2cm}J=7/2-5/2, F=3-2 	&	262.00648	 & 	12.57383  &  \ex{5.10470}{-5}	& 7\\
\hspace{0.2cm}J=5/2-3/2, F=3-2 	& 	262.06499 	 & 	12.58203 &   \ex{4.87506}{-5}	& 7\\
\hspace{0.2cm}J=5/2-3/2, F=2-1 	& 	262.06747	 &  	12.58261  &  \ex{4.45975}{-5}	 & 5\\
\midrule
\multicolumn{2}{l}{\CCH $N = 6-5$}\\ 
\hspace{0.2cm}J=13/2-11/2, F=7-6 	& 	523.97157	 & 	62.87103  &\ex{4.5697}{-4}	& 15\\
\hspace{0.2cm}J=13/2-11/2, F=6-5 	&	523.97217	 &   	62.86987 &\ex{4.5179}{-4}	& 13\\
\hspace{0.2cm}J=11/2-9/2, F=6-5 	& 	524.03390	 &    	62.88685&\ex{4.4939}{-4} 	& 13\\
\hspace{0.2cm}J=11/2-9/2, F=5-4 	& 	524.03453 	 &    	62.88757&\ex{4.4205}{-4}  	 & 11\\
\midrule
\CthHtw $J_{K_A,K_C} = 2_{1,2}-1_{0,1}$		&	85.33889		& 4.1		& \ex{2.55335}{-5}& 	15\\ 
\midrule 
\CthHtw $J_{K_A,K_C} = 4_{0,4}-3_{1,3}$		& 150.82067	&  	 19.3	& \ex{1.79933}{-4} & 9\\
\CthHtw $J_{K_A,K_C} = 4_{1,4}-3_{0,3}$		& 150.85191	&  	 17.0	& \ex{1.80086}{-4} & 27\\
\midrule
\multicolumn{2}{l}{\CthHtw $J=6-5$}	\\
\hspace{0.2cm}${K_A,K_C} = 1,6-0,5$		&	217.82215	&36.3	& \ex{5.93376}{-4}	& 	39\\ 
\hspace{0.2cm}${K_A,K_C} = 0,6-1,5$		&	217.82215	&38.6	&  \ex{5.93403}{-4}	& 	13\\ 
\midrule
\CthHtw $J_{K_A,K_C} = 5_{1,4}-4_{2,3}$		&  217.94005 	& 33.1	& \ex{4.42680 }{-4}& 	33\\ 
\midrule
\CthHtw $J_{K_A,K_C} = 4_{3,2}-3_{2,1}$		&  227.16913 	&	26.7& \ex{3.42484}{-4}& 27\\ 
\midrule
\multicolumn{2}{l}{\CthHtw $J=7-6$}	\\
\hspace{0.2cm}${K_A,K_C} = 0,7-1,6$		&  251.31434	&	48.3 & \ex{9.34953}{-4}& 	45\\ 
\hspace{0.2cm}${K_A,K_C} = 1,7-0,6$		&  251.31434	&	50.7 & \ex{9.34781}{-4}& 	15\\ 
\midrule
\multicolumn{2}{l}{\CthHtw $J=6-5$}	\\
\hspace{0.2cm}${K_A,K_C} = 2,5-1,4$		&  251.52730 	&	45.1 & \ex{7.42318}{-4}& 39\\ 

\bottomrule	
\end{tabular}\\
{\normalsize References: CH: \citep{mccarthy06} \& CDMS;  \CCH: \citep{padovani09} \& CDMS; \CthHtw: JPL}}
\end{table}

\begin{figure*}[ht]
\centering
\includegraphics[width=1\linewidth]{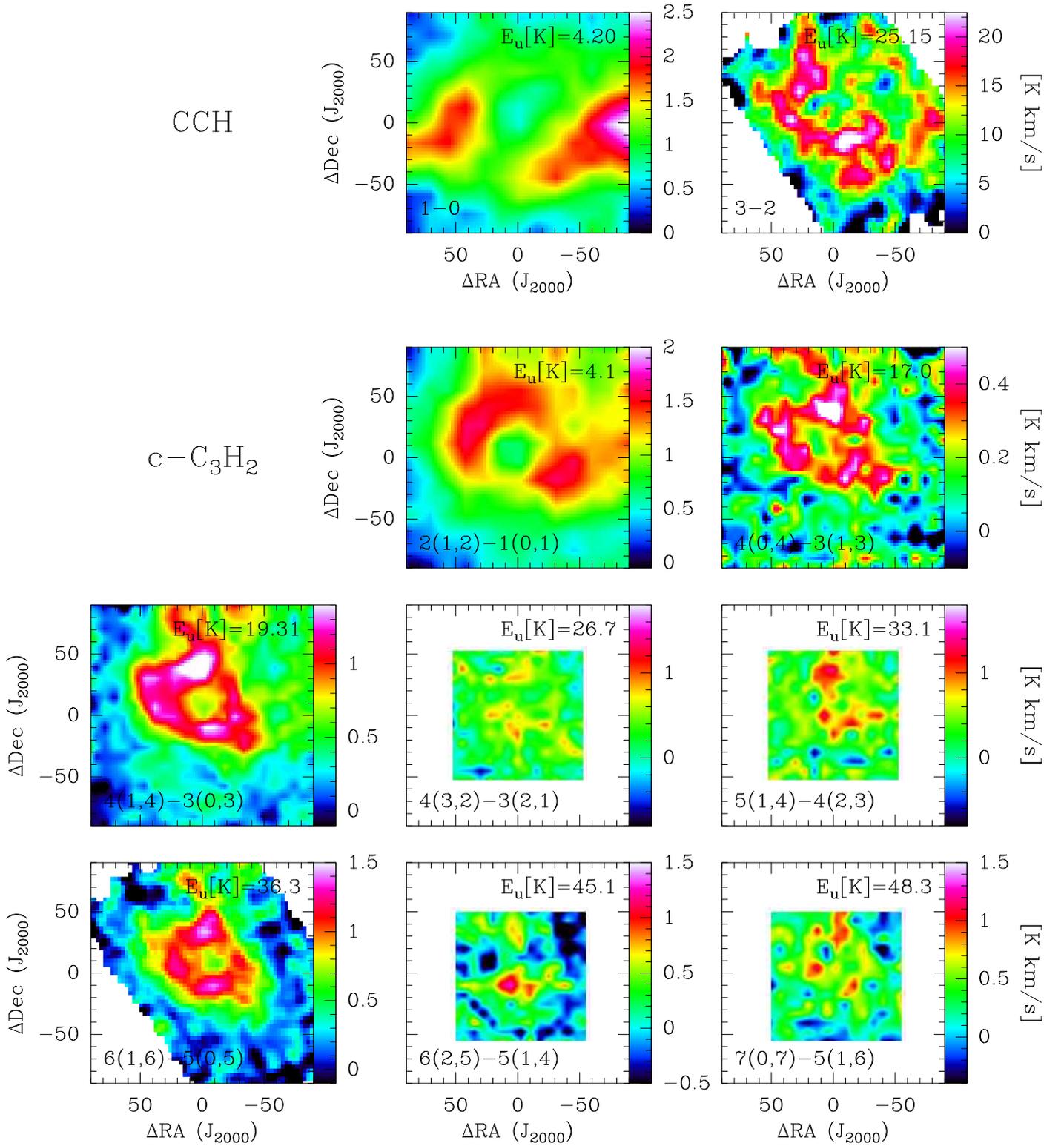}
\caption{Integrated intensity of all the observed \CthHtw and \CCH transitions. The upper state energy level and the principal quantum numbers are shown for each transition.  
The smaller-scale maps have been convolved to the spatial resolution of 15\arcsec to improve the S/N ratio.}
\label{fig_images_convolved}
\end{figure*}

\begin{figure*}
\centering
\includegraphics[width=0.3\linewidth]{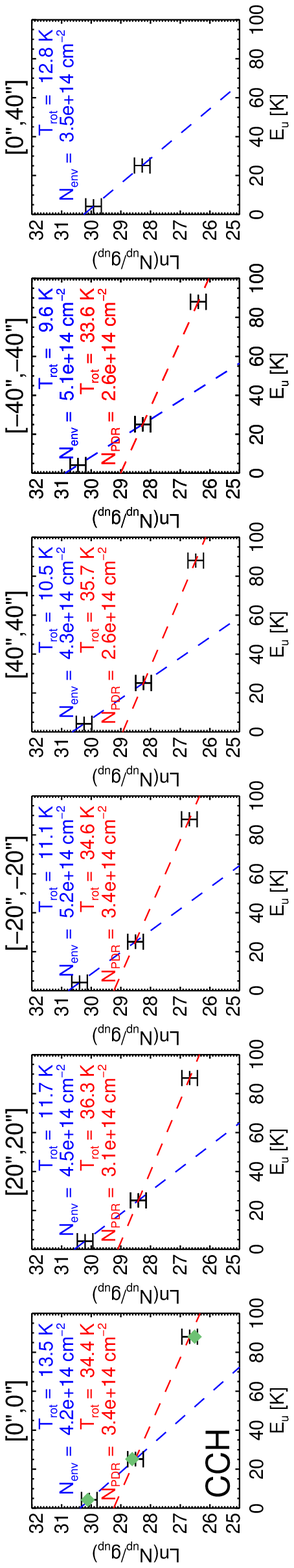}
\includegraphics[width=0.3\linewidth]{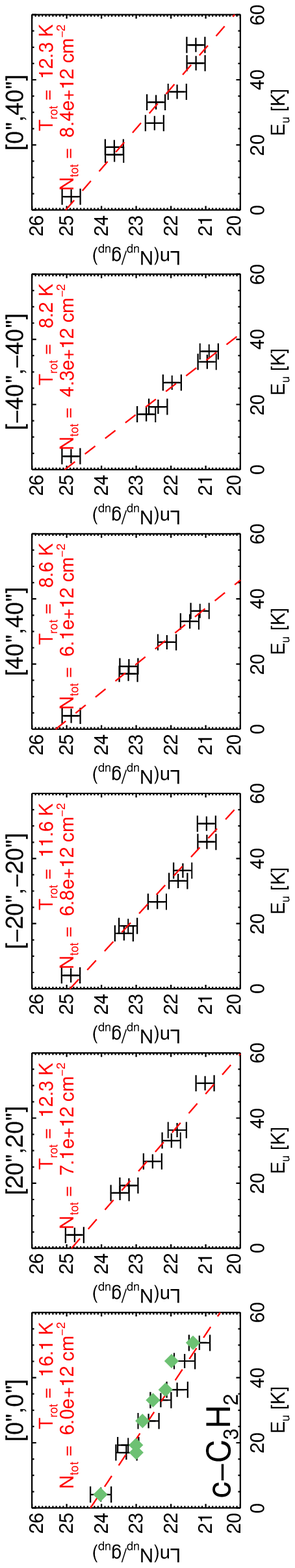}
\includegraphics[width=0.3\linewidth]{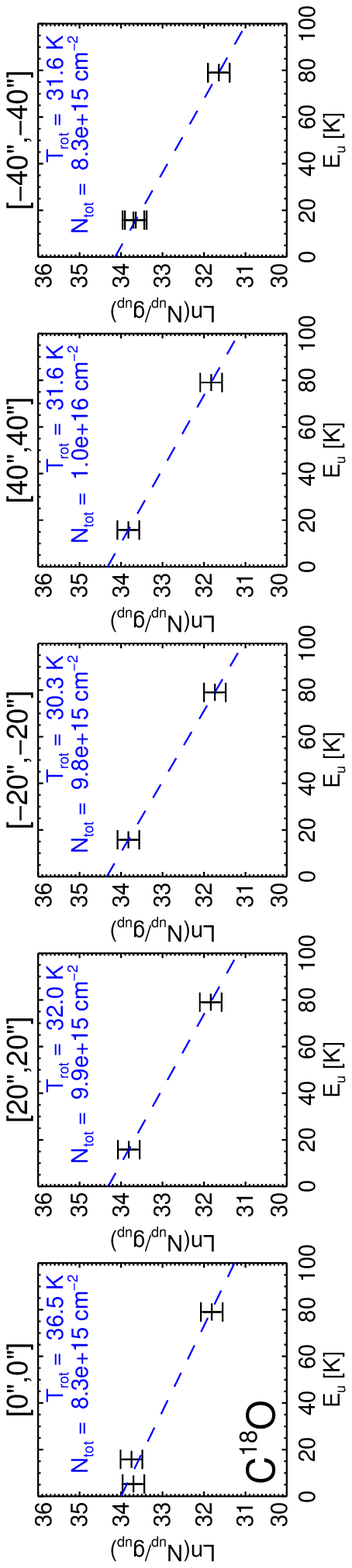}
\caption{Rotational diagrams for \CCH (top), \CthHtw (middle) and \CeiO (bottom) along the \herschel strip. Data have been convolved to 40\arcsec\, for \CCH and \CeiO, and to 29\arcsec\, for \CthHtw.  Green diamonds represent the MADEX simulation towards the IF (cf. Sect. \ref{subsec_madex})}.
\label{fig_rotdiacch}
\end{figure*}

\end{appendix}

\end{document}